\documentclass[12pt,a4paper,english,amssymb,nofootinbib,superscriptaddress]{revtex4-2}
\usepackage{lmodern}
\usepackage{lmodern}

\usepackage[T1]{fontenc}
\usepackage[latin9]{inputenc}
\usepackage{babel}
\usepackage{mathrsfs}
\usepackage{tcolorbox}
\usepackage{amsmath}
\usepackage{amssymb}
\usepackage{graphicx}
\usepackage{esint}
\usepackage[pdfusetitle,
 bookmarks=true,bookmarksnumbered=false,bookmarksopen=false,
 breaklinks=false,pdfborder={0 0 1},backref=false,colorlinks=false]
 {hyperref}

\makeatletter

\@ifundefined{textcolor}{}
{%
 \definecolor{BLACK}{gray}{0}
 \definecolor{WHITE}{gray}{1}
 \definecolor{RED}{rgb}{1,0,0}
 \definecolor{GREEN}{rgb}{0,1,0}
 \definecolor{BLUE}{rgb}{0,0,1}
 \definecolor{CYAN}{cmyk}{1,0,0,0}
 \definecolor{MAGENTA}{cmyk}{0,1,0,0}
 \definecolor{YELLOW}{cmyk}{0,0,1,0}
}

\usepackage{babel}

\def\be{\begin{equation}}
\def\ee{\end{equation}}

\@ifundefined{textcolor}{}{%
	\definecolor{BLACK}{gray}{0}
	\definecolor{WHITE}{gray}{1}
	\definecolor{RED}{rgb}{1,0,0}
	\definecolor{GREEN}{rgb}{0,1,0}
	\definecolor{BLUE}{rgb}{0,0,1}
	\definecolor{CYAN}{cmyk}{1,0,0,0}
	\definecolor{MAGENTA}{cmyk}{0,1,0,0}
	\definecolor{YELLOW}{cmyk}{0,0,1,0}
}

\usepackage{latexsym}\usepackage{bm}

\makeatother

\begin{document}
\title{Constraints and Time Evolution in Generic $f$(Riemann) Gravity}
\author{Emel Altas}
\email{emelaltas@kmu.edu.tr}

\affiliation{Department of Physics,\\
 Karamanoglu Mehmetbey University, 70100, Karaman, Turkey}
\author{Bayram Tekin\footnote{Corresponding author} }
\email{btekin@metu.edu.tr}

\affiliation{Department of Physics,\\
 Middle East Technical University, 06800, Ankara, Turkey}
\date{\today}
\begin{abstract}
\noindent We give a detailed canonical analysis of the $n$-dimensional $f$(Riemann)
gravity, correcting the earlier results in the literature. We also
write the field equations in the Fischer-Marsden form which is amenable to identifying the non-stationary energy on a spacelike hypersurface.
We give pure $R^{2}$ and $R_{\mu\nu}R^{\mu\nu}$ theories as examples. 
\end{abstract}
\maketitle

\section{Introduction}

Within the paradigm of effective field theories, General Relativity augmented with dark matter and dark energy is the lowest-order effective theory of gravity that works remarkably well at small and large scales.  Of course, it is expected to be modified at extremely high energies, or extremely short distances, for example around the black hole or Big Bang singularities. The modifications can be computed from a microscopic theory, such as string theory, alternatively, one can study generic modifications consistent with symmetries (see, for example, \cite {Petrov} for more arguments of the raison d'\^{e}tre of modified gravity theories). A large subclass of modified gravity theories is described by the action of the form\footnote{This manuscript is written for the volume ``Fields, Gravity, Strings and Beyond: In Memory of Stanley Deser'' edited by M. Henneaux, R.
I. Nepomechie, and D. Seminara. Deser (1931-2023) was very interested in modified theories of gravity: we dedicate this work to him. For personal reminiscence, the reader is invited to read \cite{Deser_Tekin}. } 
\begin{equation}
S=\frac{1}{2}\intop_{\mathcal{M}}d\thinspace^{n}x~\sqrt{-g}~f(R_{\mu\nu\rho\sigma}),\label{action1}
\end{equation}
where $f(R_{\mu\nu\rho\sigma})$ is a differentiable scalar invariant of the Riemann curvature tensor (we are taking the Newton's constant
$\kappa=1$). We consider the metric tensor to be the independent field. We shall work in the metric formulation and in $n$ dimensions.
Canonical analysis of this type of theory was given in the pioneering work \cite{Deruelle}, whose notation we shall adapt here. Our analysis mostly agrees with those of \cite{Deruelle}, however, we shall make some corrections and also recast the Hamiltonian formulation of the theory in the rather beautiful Fischer-Marsden form \cite{Fischer-Marsden}.
 Such a construction easily allows one to define the Killing initial (KID) and the approximate KIDs that are used in the definition of non-stationary energy contained in a spatial hypersurface \cite{Dain,Kroon1,Altas1,Kroon2,Altas2}.
 These computations would be relevant to identifying the initial gravitational wave content. Many papers are dedicated to various aspects of the $f(R_{\mu\nu\rho\sigma})$ theory. For example in \cite{TekinRapid},
 the particle spectrum, the masses of the perturbative excitations of this generic theory with one massive spin-$2$, one massless spin-$2$,
 and a massive spin-$0$ particle was given around any one of its constant curvature vacua. In \cite{Senturk}, conserved charges of the theory, such as energy-momentum and angular momentum, were constructed.

Here one of our goals is to expound upon the Arnowitt-Deser-Misner
(ADM) analysis \cite{ADM} and give sufficient details of the computations, so that the reader can follow all the details rather easily. We also
apply our results to pure $R^{2}$ and $R_{\mu\nu}R^{\mu\nu}$ theories.
We made a meticulous effort to write all the proofs of our statements in the appendices, so as not to significantly cut the flow of the discussion in the main text.

The layout of this paper is as follows: In section II, we introduce the action and the field equations of $f$(Riemann) theories. In section III, we summarize the ADM splitting \cite{ADM} of the action, which yields the Hamiltonian, the constraint equations, and then the time-evolution equations of the initial data on the hypersurface. In section IV, we introduce the construction of the nonstationary energy for
$f$(Riemann) theories. In section V, we consider the $R^{2}$ and
$R_{\mu\nu}R^{\mu\nu}$ theories as examples. The computations are long, therefore, we give most of the details of the calculations in Appendices.

\section{Field Equations of $f$(Riemann) theories}

The field equations coming from variation of the action (\ref{action1})
in the presence of a minimally coupled matter field are\footnote{There is a sign difference in the second term of the field equations
in \cite{Deruelle}.} 
\begin{equation}
-\frac{1}{2}g^{\mu\nu}f-R_{~\ \gamma\rho\sigma}^{(\mu}\frac{\partial f}{\partial R_{\nu)\gamma\rho\sigma}}-2\nabla_{\sigma}\nabla_{\rho}\frac{\partial f}{\partial R_{\sigma(\mu\nu)\rho}}=T^{\mu\nu},\label{EOM1}
\end{equation}
where the round brackets denote symmetrization with a factor of $1/2$.
In Appendix A, we gave the ADM decompositions of the necessary spacetime
tensor fields, and then the proof of (\ref{EOM1}) is given in Appendix
B. This equation includes fourth-order derivatives of the metric.
One introduces auxiliary variables to simplify the ensuing discussion
and lower the number of derivatives. Following \cite{Deruelle}, let
us consider the "mother action":

\begin{tcolorbox}
\begin{equation}
S=\frac{1}{2}\intop_{\mathcal{M}}d\thinspace^{n}x~\sqrt{-g}~\biggl(f(\rho_{\mu\nu\rho\sigma})+\varphi^{\mu\nu\rho\sigma}(R_{\mu\nu\rho\sigma}-\rho_{\mu\nu\rho\sigma})\biggr),\label{action}
\end{equation}
\end{tcolorbox}
\vphantom{}\\
where two auxiliary fields $(\rho,\varphi)$ have been introduced.
These rank-4 tensors have the same symmetries as the Riemann tensor
and are assumed to be independent of each other and the metric $g_{\mu\nu}$
tensor. Therefore, variation of the action with respect to all fields
can be written as 
\begin{align}
\delta S & =\frac{1}{2}\intop_{\mathcal{M}}d\thinspace^{n}x~\Biggl(\delta\sqrt{-g}~\Bigl(f(\rho_{\mu\nu\rho\sigma})+\varphi^{\mu\nu\rho\sigma}(R_{\mu\nu\rho\sigma}-\rho_{\mu\nu\rho\sigma})\Bigr)\\
 & +\sqrt{-g}~\Bigl(\delta f(\rho_{\mu\nu\rho\sigma})+\delta\varphi^{\mu\nu\rho\sigma}(R_{\mu\nu\rho\sigma}-\rho_{\mu\nu\rho\sigma})+\varphi^{\mu\nu\rho\sigma}(\delta R_{\mu\nu\rho\sigma}-\delta\rho_{\mu\nu\rho\sigma})\Bigr)\Biggr).\nonumber 
\end{align}
One has $\delta f(\rho_{\mu\nu\rho\sigma})=\frac{\partial f}{\partial\rho_{\mu\nu\rho\sigma}}\delta\rho_{\mu\nu\rho\sigma}$,
and the variation of the Riemann tensor reads 
\begin{align}
\delta R_{\mu\nu\rho\sigma} & =\delta g_{\mu\lambda}R_{~~\nu\rho\sigma}^{\lambda}+g_{\mu\lambda}\delta R_{~~\nu\rho\sigma}^{\lambda},\nonumber \\
 & =\delta g_{\mu\lambda}R_{~~\nu\rho\sigma}^{\lambda}+\nabla_{\rho}\nabla_{[\nu}\delta g_{\mu]\sigma}+\nabla_{[\rho}\nabla_{\sigma]}\delta g_{\mu\nu}+\nabla_{\sigma}\nabla_{[\mu}\delta g_{\nu]\rho}.
\end{align}
Due to the symmetries of the tensor fields, one gets 
\begin{equation}
\varphi^{\mu\nu\rho\sigma}(\delta R_{\mu\nu\rho\sigma}-\delta\rho_{\mu\nu\rho\sigma})=\delta g_{\mu\nu}R_{~~\lambda\rho\sigma}^{(\mu}\varphi^{\nu)\lambda\rho\sigma}+2\varphi^{\sigma(\mu\nu)\rho}\nabla_{\rho}\nabla_{\sigma}\delta g_{\mu\nu}-\varphi^{\mu\nu\rho\sigma}\delta\rho_{\mu\nu\rho\sigma}.
\end{equation}
Using integration by parts and defining the following tensor field\footnote{Note that in equation (2.3) of \cite{Deruelle} there is an additional
term involving the derivative of $f$ with respect to $\rho$, which
should not exist.} 
\begin{equation}
\mathcal{E^{\mu\nu}}:=-R_{~~\lambda\rho\sigma}^{(\mu}\varphi^{\nu)\lambda\rho\sigma}-2\nabla_{\sigma}\nabla_{\rho}\varphi^{\sigma(\mu\nu)\rho}-\frac{1}{2}g^{\mu\nu}\Bigl(f(\rho_{\lambda\gamma\rho\sigma})+\varphi^{\lambda\gamma\rho\sigma}(R_{\lambda\gamma\rho\sigma}-\rho_{\lambda\gamma\rho\sigma})\Bigr),
\end{equation}
one can express the variation of the total action as 
\begin{equation}
\delta S=\frac{1}{2}\intop_{\mathcal{M}}d\thinspace^{n}x~\sqrt{-g}\Biggl(-\mathcal{E^{\mu\nu}}\delta g_{\mu\nu}+\delta\varphi^{\mu\nu\rho\sigma}(R_{\mu\nu\rho\sigma}-\rho_{\mu\nu\rho\sigma})+\left(\frac{\partial f}{\partial\rho_{\mu\nu\rho\sigma}}-\varphi^{\mu\nu\rho\sigma}\right)\delta\rho_{\mu\nu\rho\sigma}\Biggr).\label{variation=000020of=000020the=000020action}
\end{equation}
In Appendix C, one can see more details of this computation including
the variation of the "mother action". From the last expression,
one gets a set of field equations. 
\begin{itemize}
\item Variation with respect to $\varphi^{\mu\nu\rho\sigma}$ gives 
\begin{equation}
R_{\mu\nu\rho\sigma}=\rho_{\mu\nu\rho\sigma}.
\end{equation}
\item Variation with respect to $\rho_{\mu\nu\rho\sigma}$ yields 
\begin{equation}
\varphi^{\mu\nu\rho\sigma}=\frac{\partial f}{\partial\rho_{\mu\nu\rho\sigma}}=\frac{\partial f}{\partial R_{\mu\nu\rho\sigma}},
\end{equation}
where for the second equality we have used the previous result. 
\item The metric variation in the action yields 
\begin{equation}
\mathcal{E^{\mu\nu}}=T^{\mu\nu}.
\end{equation}
\end{itemize}
\noindent Substituting the additional equations in the explicit form
of $\mathcal{E^{\mu\nu}}$, we recover the original fourth order derivative
equations (\ref{EOM1}). We assumed minimal coupling of matter and
the metric, and no direct coupling of the matter to the auxiliary
fields $\rho$ and $\varphi$.

\section{ADM DECOMPOSITION of the $f$(Riemann) Theory }

Let us assume that the topology of the spacetime manifold is $\mathscr{M}=\mathbb{R}\times\Sigma$,
with the first factor being the time dimension and $\Sigma$ being
a spacelike hypersurface. In Einstein's gravity, the initial data
constitute the Riemannian metric $\gamma$ and the extrinsic curvature
$K$ on $\Sigma$ together with the initial matter. The connection
on the hypersurface satisfies the metric compatibility condition:
${D}_{i}\gamma_{jk}=0$. See Fig. 1 for the slicing of the spacetime.

\begin{figure}
~~\includegraphics[scale=0.4]{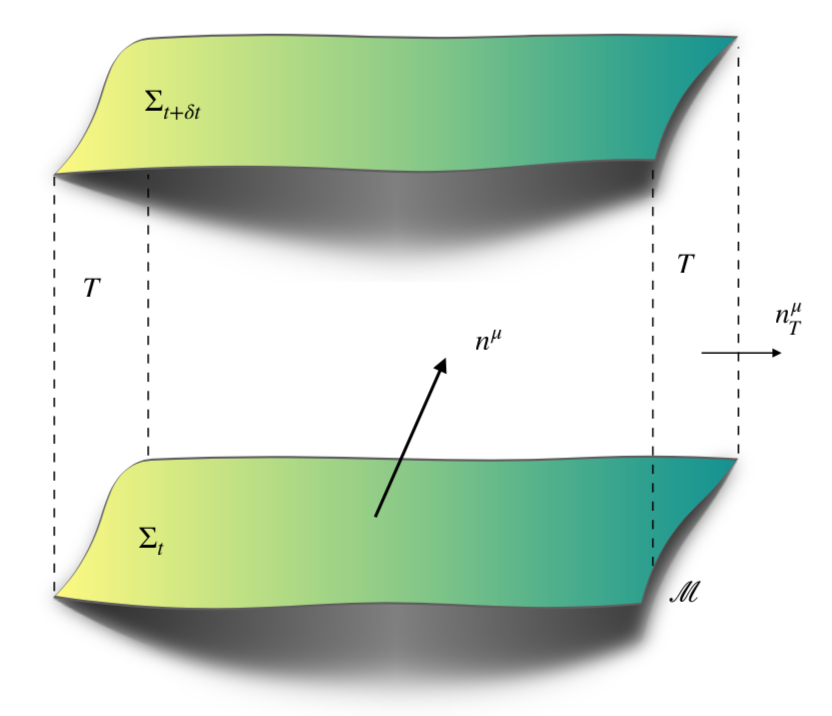}

\caption{Slicing of the spacetime in terms of co-dimension one spatial hypersurface.}
\label{fig1} 
\end{figure}

The metric in terms of the lapse function and the shift vector reads
\begin{equation}
ds^{2}=\left(N_{i}N^{i}-N^{2}\right)dt^{2}+2N_{i}~dt\thinspace dx^{i}+\gamma_{ij}\thinspace dx^{i}\thinspace dx^{j}.\label{ADMdecompositionofmetric}
\end{equation}
More explicitly, the components of the metric and the inverse metric
are 
\begin{equation}
g_{00}=-N^{2}+N_{i}N^{i},~\ ~~~~\ ~\ ~\ ~\ g_{0i}=N_{i},~~~~~\ \ g_{ij}=\gamma_{ij},
\end{equation}
\begin{equation}
g^{00}=-\frac{1}{N^{2}},~~~~g^{0i}=\frac{1}{N^{2}}N^{i},~~~~g^{ij}=\gamma^{ij}-\frac{1}{N^{2}}N^{i}N^{j}.
\end{equation}
We choose the future-pointing unit normal vector $n^{\mu}$ as 
\begin{equation}
n^{\mu}=\left(\frac{1}{N},-\frac{N^{i}}{N}\right),~~~~~~~~~~~~~~~~n_{\mu}=(-N,\vec{0}),
\end{equation}
while the extrinsic curvature in terms of the unit-normal vector reads
as 
\begin{equation}
K_{ij}=\nabla_{i}n_{j}=\frac{1}{2N}\left(\dot{\gamma}_{ij}-D_{i}N_{j}-D_{j}N_{i}\right),
\end{equation}
where $\dot{\gamma}_{ij}:=\partial_{0}\gamma_{ij}$. To rewrite the
action (\ref{action}) in terms of the ADM fields, we start with the
equality 
\begin{equation}
\varphi^{\mu\nu\rho\sigma}(R_{\mu\nu\rho\sigma}-\rho_{\mu\nu\rho\sigma})=\varphi^{ijkl}(R_{ijkl}-\rho_{ijkl})+4\varphi^{ijk0}(R_{ijk0}-\rho_{ijk0})+4\varphi^{i0j0}(R_{i0j0}-\rho_{i0j0}).
\end{equation}
We introduce the ADM splitting of the corresponding components of
the Riemann tensor in Appendix A in detail. Using these results, one
gets 
\begin{align}
\varphi^{\mu\nu\rho\sigma}(R_{\mu\nu\rho\sigma}-\rho_{\mu\nu\rho\sigma}) & =\varphi^{ijkl}(R_{ijkl}-\rho_{ijkl})+4\varphi^{ijk0}\bigl(N\left(D_{i}K_{jk}-D_{j}K_{ik}\right)+N^{l}R_{ijkl}-\rho_{ijk0}\bigr)\nonumber \\
 & +4\varphi^{i0j0}\biggl(N^{k}N^{l}\thinspace R\thinspace_{ikjl}+N^{2}K_{ik}K_{j}^{k}+NN^{k}\left(D_{i}K_{jk}+D_{j}K_{ik}-2D_{k}K_{ij}\right)\nonumber \\
 & +N(D_{i}D_{j}N-\dot{K}_{ij}+\mathcal{L_{N}}K_{ij})-\rho_{i0j0}\biggr).\label{contraction}
\end{align}
Here $\mathcal{L_{N}}$ denotes the Lie derivative along the vector
field $N^{i}$, which is defined as 
\begin{equation}
\mathcal{L_{N}}K_{ij}=N^{k}D_{k}K_{ij}+K_{ki}D_{j}N^{k}+K_{kj}D_{i}N^{k}.
\end{equation}
Following \cite{Deruelle}, let us introduce the once and twice hypersurface-projected
spatial tensor fields, respectively as 
\begin{equation}
^{\Sigma}\varphi^{ijk}:=\gamma^{il}\gamma^{jm}\gamma^{kn}n^{\mu}\varphi_{lmn\mu}\equiv\varphi^{ijk},
\end{equation}
\begin{equation}
^{\Sigma}\psi^{ij}:=-2\gamma^{ik}\gamma^{jl}n^{\mu}n^{\nu}\varphi_{k\mu l\nu}\equiv\psi^{ij},
\end{equation}
which can equivalently be written as 
\begin{eqnarray}
 &  & \varphi^{ijk}=-N\varphi^{ijk0},\\
 &  & \psi^{ij}=-2N^{2}\varphi^{i0j0}.
\end{eqnarray}
Similarly, for the other auxiliary field, we introduce 
\begin{eqnarray}
 &  & ^{\Sigma}\rho_{ijk}:=n^{\mu}\rho_{ijk\mu}\equiv\rho_{ijk},\\
 &  & ^{\Sigma}\Omega_{ij}:=n^{\mu}n^{\nu}\rho_{i\mu j\nu}\equiv\Omega_{ij}.
\end{eqnarray}
The last two expressions explicitly read 
\begin{eqnarray}
 &  & \rho_{ijk}=\frac{1}{N}\rho_{ijk0}-\frac{N^{l}}{N}\rho_{ijkl},\\
 &  & \Omega_{ij}=\frac{1}{N^{2}}\rho_{i0j0}-\frac{N^{k}N^{l}}{N^{2}}\rho_{jkil}-\frac{N^{k}}{N}\left(\rho_{ikj}+\rho_{jki}\right),
\end{eqnarray}
where $\rho_{ijkl}$ and $\varphi^{ijkl}$ themselves are spatial
tensor fields on the hypersurface by assumption: namely, one has $\rho_{ijkl}={}^{\Sigma}\rho_{ijkl}$
and $\varphi^{ijkl}={}^{\Sigma}\varphi^{ijkl}$, and for the sake
of brevity, we dropped the index $\Sigma$. In other words, we can
express the spacetime tensor components $\rho_{ijk0}$ and $\rho_{i0j0}$
in terms of purely spatial tensor fields: 
\begin{eqnarray}\label{28}
 &  & \rho_{ijk0}=N\thinspace\rho_{ijk}+N^{l}\thinspace\rho_{ijkl},\nonumber \\
 &  & \rho_{i0j0}=N^{2}~\Omega_{ij}+N^{k}N^{l}~\rho_{ikjl}+NN^{k}\left(\rho_{ikj}+\rho_{jki}\right).
\end{eqnarray}
So then, the contracted term (\ref{contraction}) becomes\footnote{In Appendix D, we give more details of this computation.}
\begin{align}
\varphi^{\mu\nu\rho\sigma}(R_{\mu\nu\rho\sigma}-\rho_{\mu\nu\rho\sigma}) & =\varphi^{ijkl}(R_{ijkl}-\rho_{ijkl})\label{contraction-1}\\
 & -4\varphi^{ijk}\biggl(D_{i}K_{jk}-D_{j}K_{ik}-\rho_{ijk}+\frac{N^{l}}{N}(R_{ijkl}-\rho_{ijkl})\biggr)\nonumber \\
 & -2\psi^{ij}\biggl(\frac{N^{k}N^{l}}{N^{2}}(R\thinspace_{ikjl}-\rho_{ikjl})+\frac{1}{N}\left(D_{i}D_{j}N-\dot{K}_{ij}+\mathcal{L_{N}}K_{ij}\right)\nonumber \\
 & +K_{ik}K_{j}^{k}-\Omega_{ij}+\frac{N^{k}}{N}\left(D_{i}K_{jk}+D_{j}K_{ik}-2D_{k}K_{ij}-\rho_{ikj}-\rho_{jki}\right)\biggr).\nonumber 
\end{align}
Inserting all of the variations to the generalized action (\ref{action}),
one arrives at 
\begin{align}
S & =\intop_{\mathcal{M}}d\thinspace^{n}x~N\sqrt{\gamma}~\Biggl(\frac{f(\rho,\Omega)}{2}+\frac{1}{2}\varphi^{ijkl}(R_{ijkl}-\rho_{ijkl})\label{red}\\
 & -2\varphi^{ijk}\biggl(D_{i}K_{jk}-D_{j}K_{ik}-\rho_{ijk}+\frac{N^{l}}{N}(R_{ijkl}-\rho_{ijkl})\biggr)\nonumber \\
 & -\psi^{ij}\biggl(\frac{N^{k}N^{l}}{N^{2}}(R\thinspace_{ikjl}-\rho_{ikjl})+\frac{1}{N}\left(D_{i}D_{j}N-\dot{K}_{ij}+\mathcal{L_{N}}K_{ij}\right)\nonumber \\
 & +K_{ik}K_{j}^{k}-\Omega_{ij}+\frac{N^{k}}{N}\left(D_{i}K_{jk}+D_{j}K_{ik}-2D_{k}K_{ij}-\rho_{ikj}-\rho_{jki}\right)\biggr)\Biggr).\nonumber 
\end{align}
Now let us eliminate the spatial tensors $\varphi^{ijkl}$ and $\varphi^{ijk}$,
which are nondynamical. The field equation for $\varphi^{ijkl}$ is
\begin{equation}
\rho_{ijkl}=R_{ijkl}=\thinspace^{\Sigma}R_{ijkl}+K_{ik}K_{jl}-K_{il}K_{jk},\label{constraint=000020rho1}
\end{equation}
while the field equation for $\varphi^{ijk}$ is 
\begin{equation}
D_{i}K_{jk}-D_{j}K_{ik}-\rho_{ijk}+\frac{N^{l}}{N}(R_{ijkl}-\rho_{ijkl})=0.
\end{equation}
The last two terms inside the round brackets cancel each other due
to the constraint and one arrives at the second constraint 
\begin{equation}
\rho_{ijk}=D_{i}K_{jk}-D_{j}K_{ik}.\label{constraint=000020rho2}
\end{equation}
Using the constraints, the action (\ref{red}) reduces to 
\begin{equation}
S=\intop_{\mathcal{M}}d\thinspace^{n}x~N\sqrt{\gamma}~\biggl(\frac{f(\rho,\Omega)}{2}-\psi^{ij}\left(K_{ik}K_{j}^{k}-\Omega_{ij}+\frac{1}{N}D_{i}D_{j}N\right)+\frac{1}{N}\left(\psi^{ij}\dot{K}_{ij}-\psi^{ij}\mathcal{L_{N}}K_{ij}\right)\biggr).
\end{equation}
Using integration by parts, one can rewrite the last two terms, ignoring
the boundary contributions, to arrive at the action $S=\intop_{\mathcal{M}}d\thinspace^{n}x~\mathcal{L}$,
where the Lagrangian density is

\begin{tcolorbox}
\begin{equation}
\mathcal{L}=N\sqrt{\gamma}~\Biggl(\frac{f(\rho,\Omega)}{2}-\psi^{ij}\left(KK_{ij}+K_{ik}K_{j}^{k}-\Omega_{ij}+\frac{1}{N}D_{i}D_{j}N\right)-\frac{1}{N}K_{ij}\left(\dot{\psi}^{ij}-\mathcal{L_{N}}\psi^{ij}\right)\Biggr).\label{lagrangian}
\end{equation}
\end{tcolorbox}
\hphantom{}\\
Therefore, with the help of the auxiliary fields, we have managed
to recast the higher derivative action as a lower derivative one as
desired. It is now easier to find the Hamiltonian of the theory from
this lower-derivative action.

\subsection{Hamiltonian of the theory}

\noindent The dynamical fields are ($\gamma_{ij},$ $\psi^{ij}$),
and hence one needs to introduce the two canonical momenta corresponding
to these dynamical fields. These are 
\begin{equation}
\Pi_{ij}:=\frac{\delta\mathcal{L}}{\delta\partial_{0}\psi^{ij}},
\end{equation}
and 
\begin{equation}
p^{ij}:=\frac{\delta\mathcal{L}}{\delta\partial_{0}\gamma_{ij}}=\frac{\delta\mathcal{L}}{\delta K_{lm}}\frac{\delta K_{lm}}{\delta\dot{\gamma}_{ij}}.
\end{equation}
After a straightforward computation, one ends up with 
\begin{equation}
\Pi_{ij}=-\sqrt{\gamma}K_{ij}.\label{pirelationk}
\end{equation}
The constraints (\ref{constraint=000020rho1}, \ref{constraint=000020rho2})
can now be recast in terms of the canonical momenta as 
\begin{equation}
\rho_{ijk}=D_{j}\left(\frac{\Pi_{ik}}{\sqrt{\gamma}}\right)-D_{i}\left(\frac{\Pi_{jk}}{\sqrt{\gamma}}\right),
\end{equation}
\begin{equation}
\rho_{ijkl}=R_{ijkl}=\thinspace^{\Sigma}R_{ijkl}+\frac{1}{\gamma}\left(\Pi_{ik}\Pi_{jl}-\Pi_{il}\Pi_{jk}\right).
\end{equation}
Similarly one arrives at 
\begin{equation}
p^{ij}=\frac{\sqrt{\gamma}}{4}\frac{\delta f}{\delta K_{ij}}+\frac{\sqrt{\gamma}}{2}\left(\frac{1}{N}(\mathcal{L_{N}}\psi^{ij}-\dot{\psi}^{ij})-\gamma^{ij}\psi^{kl}K_{kl}-K\psi^{ij}-\psi^{ik}K_{k}^{j}-\psi^{jk}K_{k}^{i}\right).
\end{equation}
Then the Hamiltonian density 
\begin{equation}
\mathcal{H}=p^{ij}\dot{\gamma}_{ij}+\pi_{ij}\dot{\psi}^{ij}-\mathcal{L},
\end{equation}
becomes 
\begin{align}
\mathcal{H} & =2NK_{ij}\thinspace p^{ij}-\sqrt{\gamma}K_{ij}\thinspace\mathcal{L_{N}}\psi^{ij}+2p^{ij}D_{i}N_{j}\nonumber \\
 & +\sqrt{\gamma}N\left(-\frac{f}{2}+\frac{1}{N}\psi^{ij}D_{i}D_{j}N-\psi^{ij}\Omega_{ij}+K\psi^{kl}K_{kl}+K_{ij}\psi^{jk}K_{k}^{i}\right).\label{Hamiltonian}
\end{align}

\subsection{Constraint equations}

Up to a boundary contribution, one can express the Hamiltonian density
as a sum of constraint equations: 
\begin{equation}
\mathcal{H}=N\varPhi_{0}+N^{i}\varPhi_{i}.
\end{equation}
Here $\varPhi_{0}$ denotes the Hamiltonian constraint and $\varPhi_{i}$
denotes the momentum constraints. Using (\ref{Hamiltonian}), one
can rewrite the Hamiltonian density as 
\begin{align}
\mathcal{H} & =N\sqrt{\gamma}\left(\frac{2}{\sqrt{\gamma}}K_{ij}\thinspace p^{ij}-\frac{f}{2}+D_{i}D_{j}\psi^{ij}-\psi^{ij}\Omega_{ij}+K\thinspace K_{ij}\psi^{ij}+K_{ij}K_{k}^{i}\psi^{jk}\right)\nonumber \\
 & +N^{i}\sqrt{\gamma}\left(-2D_{k}\left(\frac{p_{i}^{k}}{\sqrt{\gamma}}\right)-K_{kl}D_{i}\psi^{kl}-2D_{k}(\psi^{kl}K_{li})\right).
\end{align}
Equivalently, in terms of the canonical momenta one obtains 
\begin{tcolorbox}
\begin{align}
\mathcal{H} & =N\sqrt{\gamma}\left(-\frac{f}{2}+D_{i}D_{j}\psi^{ij}-\psi^{ij}\Omega_{ij}\right)+\frac{N}{\sqrt{\gamma}}\left(-2\Pi_{ij}p^{ij}+\Pi~\Pi_{ij}\psi^{ij}+\Pi_{ij}\Pi_{k}^{i}\psi^{jk}\right)\nonumber \\
 & +N^{i}\left(-2\sqrt{\gamma}D_{k}\left(\frac{p_{i}^{k}}{\sqrt{\gamma}}\right)+\Pi_{kl}D_{i}\psi^{kl}+2\sqrt{\gamma}D_{k}\left(\psi^{kl}\frac{\Pi_{li}}{\sqrt{\gamma}}\right)\right).\label{Hamiltonain}
\end{align}
\end{tcolorbox}
 \hphantom{}\\
Therefore the constraints are 
\begin{eqnarray}
 &  & \varPhi_{0}=\sqrt{\gamma}\left(\frac{2}{\sqrt{\gamma}}K_{ij}\thinspace p^{ij}-\frac{f}{2}+D_{i}D_{j}\psi^{ij}-\psi^{ij}\thinspace\Omega_{ij}+K\thinspace K_{ij}\psi^{ij}+K_{ij}K_{k}^{i}\psi^{jk}\right),\\
 &  & \varPhi_{i}=\sqrt{\gamma}\left(-2D_{k}\left(\frac{p_{i}^{k}}{\sqrt{\gamma}}\right)-K_{kl}D_{i}\psi^{kl}-2D_{k}(\psi^{kl}K_{li})\right).
\end{eqnarray}
The constraints vanish in a vacuum; but if there is a non-zero energy
momentum tensor, then they must be equal to the corresponding projection
of the energy-momentum tensor onto the initial hypersurface: 
\begin{eqnarray}
\varPhi_{0}=2T_{nn} & = & \frac{2}{N^{2}}\left(2N^{i}T_{0i}-T_{00}-N^{i}N^{j}T_{ij}\right),\\
\varPhi_{i}=2T_{ni} & = & \frac{2}{N}\left(N^{j}T_{ij}-T_{0i}\right).
\end{eqnarray}

In addition, the field equations of $\Omega_{ij}$ also are constraints:
$\frac{\delta\mathcal{H}}{\delta\Omega_{ij}}=0$. Finally, let us
write all the constraints.\footnote{ Note that the reader can study Appendix E for more construction details. } 
\begin{itemize}
\item The Hamiltonian constraint:

\begin{tcolorbox}
\begin{equation}
\varPhi_{0}=\sqrt{\gamma}\left(-\frac{f}{2}+D_{i}D_{j}\psi^{ij}-\Omega_{ij}\psi^{ij}\right)+\frac{1}{\sqrt{\gamma}}\left(-2\Pi_{ij}p^{ij}+\Pi\,\Pi_{ij}\psi^{ij}+\Pi_{ij}\Pi_{k}^{i}\psi^{jk}\right)=0.\label{Hamiltonian=000020constraint}
\end{equation}
\end{tcolorbox}

\item The momentum constraint:

\begin{tcolorbox}
\begin{equation}
\varPhi_{i}=-2\sqrt{\gamma}D_{k}\left(\frac{p_{i}^{k}}{\sqrt{\gamma}}\right)+\Pi_{kl}D_{i}\psi^{kl}+2\sqrt{\gamma}D_{k}\left(\psi^{kl}\frac{\Pi_{li}}{\sqrt{\gamma}}\right)=0.\label{momentum=000020constraint}
\end{equation}
\end{tcolorbox}

\item The additional constraint of the auxiliary field:

\begin{tcolorbox}
\begin{equation}
2\psi^{ij}+\frac{\delta f}{\delta\Omega_{ij}}=0.\label{constraint=000020on=000020psi}
\end{equation}
\end{tcolorbox}

\end{itemize}

\subsection{Time evolution equations}

\subsubsection{The first set: $\dot{\gamma}_{ij}$, $\dot{\psi}^{ij}$}

From now on we are going to construct the time evolution equations.\footnote{Since the computation is rather long, we delegate some details of
this section to Appendix F.} The phase space variables are ($\gamma_{ij},$ $\psi^{ij}$, $p^{ij}$,
$\Pi_{ij}$). The canonical coordinates evolve via 
\begin{equation}
\dot{\gamma}_{ij}=\frac{\delta\mathcal{H}}{\delta p^{ij}},~~~~~~~~~~~~~~~~\dot{\psi}^{ij}=\frac{\delta\mathcal{H}}{\delta\Pi_{ij}}.
\end{equation}
The definition of extrinsic curvature leads to 
\begin{equation}
\dot{\gamma}_{ij}=2NK_{ij}+D_{i}N_{j}+D_{j}N_{i}.
\end{equation}
The relation (\ref{pirelationk}) additionally yields 
\begin{equation}
\dot{\gamma}_{ij}=-\frac{2N}{\sqrt{\gamma}}\Pi_{ij}+{\mathcal{L}}_{N}\gamma_{ij},
\end{equation}
and so we can write

\begin{tcolorbox}
\begin{equation}
\dot{\gamma}_{ij}=2NK_{ij}+D_{i}N_{j}+D_{j}N_{i}=-\frac{2N}{\sqrt{\gamma}}\Pi_{ij}+{\mathcal{L}}_{N}\gamma_{ij}.\label{evolution=000020spatial=000020metric}
\end{equation}
\end{tcolorbox}
 \hphantom{}\\
On the other hand, one has 
\begin{tcolorbox}
\begin{align}
\dot{\psi}^{ij}=\frac{N}{\sqrt{\gamma}}\left(-2p^{ij}+\gamma^{ij}\Pi_{kl}\thinspace\psi^{kl}+\Pi\psi^{ij}+\Pi_{k}^{i}\psi^{jk}+\Pi_{k}^{j}\psi^{ik}\right)+\mathcal{L_{N}}\psi^{ij}-\frac{N\sqrt{\gamma}}{2}\frac{\delta f}{\delta\Pi_{ij}}.\label{evolution=000020psi}
\end{align}
\end{tcolorbox}

\subsubsection{The second set: $\dot{p}^{ij}$, $\dot{\Pi}_{ij}$}

Next, we find the time-evolution equations for the canonical momenta:

\begin{equation}
\dot{p}^{ij}=-\frac{\delta\mathcal{H}}{\delta\gamma_{ij}},~~~~~~~~~~~~~~~~\dot{\Pi}_{ij}=-\frac{\delta\mathcal{H}}{\delta\psi^{ij}}.
\end{equation}
The second one is easier to obtain since we only focus on the variations
with respect to $\psi^{ij}$. Using the Hamiltonian (\ref{Hamiltonain}),
we have 
\begin{align}
\dot{\Pi}_{ij} & =-N\sqrt{\gamma}\frac{\delta}{\delta\psi^{ij}}\left(D_{k}D_{l}\psi^{kl}-\psi^{kl}\thinspace\Omega_{kl}\right)-\frac{N}{\sqrt{\gamma}}\frac{\delta}{\delta\psi^{ij}}\left(\Pi\thinspace\Pi_{kl}\thinspace\psi^{kl}+\Pi_{mn}\Pi_{k}^{n}\psi^{mk}\right)\nonumber \\
 & -N^{m}\frac{\delta}{\delta\psi^{ij}}\left(\Pi_{kl}D_{m}\psi^{kl}+2\sqrt{\gamma}D_{k}\left(\psi^{kl}\frac{\Pi_{lm}}{\sqrt{\gamma}}\right)\right),
\end{align}
where 
\begin{equation}
\frac{\delta\psi^{kl}}{\delta\psi^{ij}}=\frac{1}{2}\left(\delta_{i}^{k}\delta_{j}^{l}+\delta_{i}^{l}\delta_{j}^{k}\right).
\end{equation}
After ignoring the total derivative terms, we get 
\begin{equation}
N\frac{\delta}{\delta\psi^{ij}}D_{k}D_{l}\psi^{kl}=D_{i}D_{j}N,
\end{equation}
and 
\begin{equation}
N^{m}\Pi_{kl}\frac{\delta}{\delta\psi^{ij}}D_{m}\psi^{kl}=-N^{m}D_{m}\Pi_{ij}-\Pi_{ij}D_{m}N^{m},
\end{equation}
and also 
\begin{equation}
N^{m}\frac{\delta}{\delta\psi^{ij}}D_{k}\left(\psi^{kl}\frac{\Pi_{lm}}{\sqrt{\gamma}}\right)=-\frac{1}{\sqrt{\gamma}}\Pi_{m(i}D_{j)}N^{m}.
\end{equation}
Substituting all of these pieces, we end up with 
\begin{tcolorbox}
\begin{equation}
\dot{\Pi}_{ij}=\sqrt{\gamma}\left(N\Omega_{ij}-D_{i}D_{j}N\right)-\frac{N}{\sqrt{\gamma}}\left(\Pi\,\Pi_{ij}+\Pi_{ik}\Pi_{j}^{k}\right)+\sqrt{\gamma}\mathcal{L_{N}}\left(\frac{\Pi_{ij}}{\sqrt{\gamma}}\right)+\Pi_{ij}D_{k}N^{k}.\label{evolution=000020Pi1}
\end{equation}
\end{tcolorbox}
\hphantom{}\\
Similarly we can find $\dot{p}^{ij}$. Clearly, we can express 
\begin{align}
\dot{p}^{ij} & =\frac{N}{\sqrt{\gamma}}\biggl(\gamma^{ij}\left(-2\Pi_{kl}p^{kl}+\Pi\,\Pi_{kl}\psi^{kl}+\Pi_{km}\Pi_{n}^{m}\psi^{nk}\right)+\Pi^{ij}\Pi_{kl}\psi^{kl}+\Pi_{l}^{i}\Pi_{k}^{j}\psi^{kl}\biggr)\nonumber \\
 & -N\sqrt{\gamma}\frac{\delta}{\delta\gamma_{ij}}D_{k}D_{l}\psi^{kl}+\frac{N}{2}\sqrt{\gamma}\frac{\delta f}{\delta\gamma_{ij}}\nonumber \\
 & -N^{m}\frac{\delta}{\delta\gamma_{ij}}\left(-2\sqrt{\gamma}\gamma_{mn}D_{k}\left(\frac{p^{kn}}{\sqrt{\gamma}}\right)+\Pi_{kl}D_{m}\psi^{kl}+2\sqrt{\gamma}D_{k}\left(\psi^{kl}\frac{\Pi_{lm}}{\sqrt{\gamma}}\right)\right),
\end{align}
where the last two terms cancel each other because of the variation
with respect to the spatial metric. Taking into account the covariant
derivatives correctly, we have\footnote{Note that in both of these equations (\ref{evolution=000020Pi1},
\ref{evolution=000020pij}), all the terms except the last term in
each one are the same as those of \cite{Deruelle}. Those two terms
are missing in that work.} 
\begin{tcolorbox}
\begin{align}
\dot{p}^{ij} & =\frac{N}{\sqrt{\gamma}}\Biggl(\gamma^{ij}\left(\Pi\,\Pi_{kl}\psi^{kl}+\Pi_{km}\Pi_{n}^{m}\psi^{nk}-2\Pi_{kl}p^{kl}\right)+\Pi^{ij}\Pi_{kl}\psi^{kl}+\Pi_{l}^{i}\Pi_{k}^{j}\psi^{kl}\Biggr)\label{evolution=000020pij}\\
 & +\frac{\sqrt{\gamma}}{2}\Biggl(D_{k}\left(\psi^{ij}D^{k}N-2\psi^{k(i}D^{j)}N\right)+\gamma^{ij}\left(ND_{k}D_{l}\psi^{kl}-\psi^{kl}D_{k}D_{l}N\right)\Biggr)\nonumber \\
 & +\sqrt{\gamma}\mathcal{L}_{N}\left(\frac{p^{ij}}{\sqrt{\gamma}}\right)+\frac{N}{2}\sqrt{\gamma}\frac{\delta f}{\delta\gamma_{ij}}+p^{ij}D_{k}N^{k}.\nonumber 
\end{align}
\end{tcolorbox}
\hphantom{}\\
In Appendix G, we gave the construction of the constraint and time
evolution equations of General Relativity using the results we have
obtained in this and in the previous sections. 

As explained in detail in \cite{Altas1}, the Hamiltonian form of
the Einstein-Hilbert action, when extremized, leads to the Fischer-Marsden
form \cite{Fischer-Marsden} of the field equations. For the generic
$f(Riemann)$ theory studied here, one can also recast the Hamiltonian
flow in a concise form as 
\begin{tcolorbox}
\begin{equation}
\frac{d}{dt}\begin{pmatrix}\gamma\\
\psi\\
p\\
\Pi
\end{pmatrix}=J\circ{\bf D\Phi^{*}}(\gamma,\psi,\pi,\Pi)({\cal {N}}),\hskip1cmJ:=\begin{pmatrix}0 & 0 & ~0~ & 1~\\
0 & 0 & 1 & 0\\
0 & -1~ & 0 & 0\\
-1 & 0 & 0 & 0
\end{pmatrix},\label{evolution}
\end{equation}
\end{tcolorbox}
\hphantom{}\\
where the small circle represents the usual matrix product.

\noindent In this matrix equation, ${\bf D\Phi^{*}}(\gamma,\psi,\pi,\Pi)$
is the formal adjoint of the linearized constraint map ($D\Phi(\gamma,\psi,\pi,\Pi)$).
Why the adjoint map appears in the Hamiltonian flow can be understood
from the discussion in \cite{Altas1}. Here ${\cal {N}}$ is the lapse-shift
vector with components $(N,N^{i})$. Observe that there is no time-evolution
when ${\bf D\Phi^{*}}(\gamma,\psi,\pi,\Pi)({\cal {N}})=0$, and these
points in the space of initial data yield Killing vectors in spacetime
\cite{Moncrief,Beig}. Such a description of Killing symmetries is
extremely useful in understanding the amount of non-stationary energy
contained in a given initial data. Here is how: if ${\cal {N}}=\xi$
is a Killing vector, say a stationary Killing vector, then the time
evolution is trivial. The failure of ${\cal {N}}$ to be a Killing
vector field is given as 
\begin{equation}
{\bf D\Phi^{*}}(\gamma,\psi,\pi,\Pi)({\cal {N}})=J^{-1}\circ\frac{d}{dt}\begin{pmatrix}\gamma\\
\psi\\
p\\
\Pi
\end{pmatrix}.
\end{equation}
Next, we discuss the non-stationary energy in this generic theory based on the approximate Killing initial data.

\section{NON-STATIONARY ENERGY IN $f$(Riemann) THEORIES}

Dain \cite{Dain} introduced the concept of non-stationary energy
 for the time-symmetric initial data in General Relativity for vacuum asymptotically flat spacetimes. That definition is based on the notion of approximate Killing initial data (KID), which is to be defined below. Dain's invariant was extended to the time-asymmetric case in
\cite{Kroon1}, and for asymptotically non-flat spacetimes in \cite{Altas1},
 where another definition based on the time-evolution equations was given. In \cite{Altas2} the construction was extended to non-vacuum spacetimes.

Let us briefly recap Dain's construction as it is not widely known and involves several subtle steps. Let the constraint covector be
${\bf \Phi}:=(\Phi_{0},\Phi_{i})$ and ${\bf D}{\bf \Phi}$ be its
linearization about a given solution initial solution. Then, ${\bf D\Phi}^{*}$
is the formal adjoint of the linearized constraint map that acts on the lapse and shift vector. A crucial tool in the construction of Dain's invariant is Bartnik's operator ${\cal {P}}$ defined as
\cite{Bartnik} 
\begin{equation}
{\cal {P}}:={\bf D\Phi}\circ\begin{pmatrix}1 & 0\\
0 & -D^{m}
\end{pmatrix},
\end{equation}
of which the formal adjoint is 
\begin{equation}
{\cal {P}}^{*}({\cal {N}}):=\begin{pmatrix}1 & 0\\
0 & D_{m}
\end{pmatrix}\circ{\bf D\Phi^{*}}({\cal {N}}).
\end{equation}
If one uses the densitized versions of the constraints, one must also
rescale the Bartnik's operator as 
\begin{equation}
\widetilde{{\cal {P}}}^{*}({\cal {N}}):=\begin{pmatrix}\gamma^{-1/2} & 0\\
0 & 1
\end{pmatrix}\circ{\cal {P}}^{*}({\cal {N}}).\label{formaladjoint_ptilde}
\end{equation}
Finally, we can write the Dain's invariant, $\mathscr{I}(\xi)$, that
quantifies the amount of non-stationary energy in the initial data
that solves the constraint equations: 
\begin{equation}
\mathscr{I}(\xi):=\intop_{\Sigma}dV~{\cal {P}}^{*}(\xi)\cdot{\cal {P}}^{*}(\xi),\label{integralinvariant}
\end{equation}
where $\xi:=(N,N^{i})$, and ${P}^{*}(\xi):={P}^{*}\begin{pmatrix}N\\
N^{k}
\end{pmatrix}$. More explicitly, in (\ref{integralinvariant}) one has 
\begin{equation}
\begin{pmatrix}N\\
N^{i}
\end{pmatrix}\cdot\begin{pmatrix}A\\
B_{i}
\end{pmatrix}:=NA+N^{i}B_{i}.
\end{equation}
The important step here is the following: in the integral (\ref{integralinvariant}),
 one considers only the lapse and shift functions that satisfy a fourth-order partial differential equation that arises in the integration
by parts as 
\begin{equation}
{\cal {P}}\circ{\cal {P}}^{*}\left(\xi\right)=0.\label{approx_KID}
\end{equation}
This is called (by Dain) the "approximate KID equation", which admit all the Killing initial data as solutions, but has more solutions than the Killing initial data.

Our formulation \cite{Altas1,Altas2} of Dain's invariant directly involves the time evolution equations since one can write the formal-adjoint
of Bartnik's operator as 
\begin{equation}
{\cal {P}}^{*}({\cal {N}}):=\begin{pmatrix}1 & 0\\
0 & D_{m}
\end{pmatrix}\circ{\bf D\Phi^{*}}({\cal {N}})=\begin{pmatrix}~1~\, & ~0\,~ & 0 & 0\\
0 & 1 & 0 & 0\\
0 & 0 & D_{m} & 0\\
0 & 0 & 0 & D_{m}
\end{pmatrix}\circ J^{-1}\circ\frac{d}{dt}\begin{pmatrix}\gamma\\
\psi\\
p\\
\Pi
\end{pmatrix}.
\end{equation}
Then, Dain's invariant for generic $f$(Riemann) theories in the time evolution formulation reads as 
\begin{equation}
\text{\ensuremath{\mathscr{I}}}({\cal {N}})=\intop_{\Sigma}dV\thinspace\left(|D_{m}\dot{\gamma}_{ij}|^{2}+|D_{m}\dot{\psi}^{ij}|^{2}+\frac{1}{\gamma}\left(|\dot{p}^{ij}|^{2}+|\dot{\Pi}_{ij}|^{2}\right)\right), \label{GeneralDain}
\end{equation}
where the time derivatives of the phase space fields appear. One must
also be careful with the notation as one has $|D_{m}\dot{\psi}^{ij}|^{2}\equiv\gamma_{ik}\gamma_{jl}D_{m}\dot{\psi}^{ij}D^{m}\dot{\psi}^{kl}$.

Let us remark on the possible use of the results of this section. 
Given initial data that solves the constraints, one can identify what fraction of that data will turn into gravitational waves using the expression (\ref{GeneralDain}). As a fully deterministic theory, this is what one expects in gravity. Unfortunately, it is generically hard to find analytical solutions to the constraints. Therefore, one needs to compute (\ref{GeneralDain}) for a numerical solution. Even in the simplest case, provided by Dain \cite{Dain} for asymptotically flat time-symmetric initial data in Einstein's theory, a numerical evaluation of the related integral that gives the non-stationary energy has not been carried out. It is an outstanding problem.\footnote{One reason this approach to the gravitational wave content of initial data has not received much attention could be the fact that Sergio Dain passed away at the age of 46 before he was able to expound upon his ideas on the topic \cite{Ash}.}

\section{Applications of The formalism }

\subsection{ The $R^{2}$ theory}

From now on, we shall adapt our results to the $R^{2}$ theory. We consider the following function 
\begin{equation}
f(\rho_{\mu\nu\rho\sigma})=\rho^{2}=g^{\mu\rho}g^{\nu\sigma}g^{\alpha\beta}g^{\gamma\kappa}\rho_{\mu\nu\rho\sigma}\rho_{\alpha\gamma\beta\kappa}
\end{equation}
to represent the $R^{2}$ theory, where $\rho=g^{\mu\rho}g^{\nu\sigma}\rho_{\mu\nu\rho\sigma}$
and so

\begin{equation}
\begin{aligned}\rho & =g^{\mu\rho}g^{v\sigma}\rho_{\mu v\rho\sigma}\\
 & =g^{v\sigma}\left(g^{0\rho}\rho_{0v\rho\sigma}+g^{i\rho}\rho_{iv\rho\sigma}\right)\\
 & =g^{v\sigma}\left(g^{00}\rho_{0v0\sigma}+g^{0i}\rho_{0vi\sigma}+g^{i0}\rho_{iv0\sigma}+g^{ij}\rho_{ivj\sigma}\right),
\end{aligned}
\end{equation}
which yields

\begin{equation}
\begin{aligned}\rho= & g^{0\sigma}\left(g^{i0}\rho_{i00\sigma}+g^{ij}\rho_{i0j\sigma}\right)+g^{k\sigma}\left(g^{00}\rho_{0k0\sigma}+g^{0i}\rho_{0ki\sigma}+g^{i0}\rho_{ik0\sigma}+g^{ij}\rho_{ikj\sigma}\right),\end{aligned}
\end{equation}
and then

\begin{equation}
\begin{aligned}\rho= & g^{00}g^{ij}\rho_{i0j0}+g^{0k}\left(g^{i0}\rho_{i00k}+g^{ij}\rho_{i0jk}\right)+g^{k0}\left(g^{0i}\rho_{0ki0}+g^{ij}\rho_{ikj0}\right)\\
 & +g^{kl}\left(g^{00}\rho_{0k0l}+g^{0i}\rho_{0kil}+g^{i0}\rho_{ik0l}+g^{ij}\rho_{ikjl}\right).
\end{aligned}
\end{equation}
The auxiliary field $\rho_{\mu v\rho\sigma}$ has the all symmetries
of the Riemann tensor. Therefore, by renaming the indices we get

\begin{equation}
\rho=g^{ij}g^{kl}\rho_{ikjl}+2\rho_{i0j0}\left(g^{00}g^{ij}-g^{0i}g^{0j}\right)+4g^{ik}g^{0j}\rho_{0ijk}.
\end{equation}
Inserting the corresponding components of the inverse spacetime metric,
we arrive at

\begin{equation}
\rho=\gamma^{ij}\gamma^{kl}\rho_{ikjl}-2\gamma^{ij}\frac{1}{N^{2}}\left(N^{k}N^{l}\rho_{ikjl}+\rho_{i0j0}-N^{k}(\rho_{0ikj}+\rho_{0jki})\right),
\end{equation}
where we have already introduced the hypersurface projected field
$\Omega_{ij}=n^{\mu}n^{v}\rho_{i\mu jv}$ with the future pointing
unit normal vector $n^{\mu}=\left(1/N_{1}-N^{i}/N\right)$. Hence we get

\[
\rho=\gamma^{ij}\gamma^{kl}\rho_{ikjl}-2\gamma^{ij}\Omega_{ij}={}^{\Sigma}\rho-2\Omega,
\]
where we have used $\rho_{ikjl}=\thinspace^{\Sigma}\rho_{ikjl}$ and
$\Omega_{ij}=\thinspace^{\Sigma}\Omega_{ij}$. Then 
\begin{equation}
f=4\gamma^{ij}\gamma^{kl}\Omega_{ij}\Omega_{kl}-4\gamma^{ij}\Omega_{ij}\gamma^{kl}\gamma^{mn}\thinspace^{\Sigma}\rho_{kmln}+\gamma^{kl}\gamma^{mn}\thinspace^{\Sigma}\rho_{kmln}\gamma^{ps}\gamma^{ij}\thinspace^{\Sigma}\rho_{pisj}.
\end{equation}
Here we have used $\gamma^{ik}\gamma^{jl}\rho_{ijkl}=\thinspace^{\Sigma}\mathbf{\rho}$,
to make the difference clear between the trace with the spacetime
metric, $\rho=g^{\mu\rho}g^{\nu\sigma}\rho_{\mu\nu\rho\sigma}$. Then,
we write 
\begin{tcolorbox}
\begin{equation}
f(\rho_{\mu\nu\rho\sigma})=\rho^{2}=\left(\thinspace^{\Sigma}\mathbf{\rho}-2\Omega\right)^{2}.\label{f_R2}
\end{equation}
\end{tcolorbox}
\hphantom{}\\
To construct the primary constraint, $\partial f/\partial\Omega_{ij}=-2\psi^{ij}$,
we need to calculate $\partial f/\partial\Omega_{ij}$. It is easy
to prove that 
\begin{equation}
\frac{\partial f}{\partial\Omega_{ij}}=8\Omega\gamma^{ij}-4\gamma^{ij}\thinspace^{\Sigma}\rho.
\end{equation}
Therefore the primary constraint of the auxiliary field is

\begin{tcolorbox}
\begin{equation}
\psi^{ij}=\gamma^{ij}\left(2\thinspace^{\Sigma}\rho-4\Omega\right).\label{psi_R2}
\end{equation}
\end{tcolorbox}
\hphantom{}\\
Recall that in General Relativity one has $\psi^{ij}=\gamma^{ij}$,
and now we have $\gamma,{}^{\Sigma}\rho,\Omega$ dependence in the
hypersurface field $\psi^{ij}$. The Hamiltonian constraint (\ref{Hamiltonian=000020constraint})
reduces to

\begin{tcolorbox}
\begin{equation}
\varPhi_{0}=2\sqrt{\gamma}\left(D_{k}D^{k}\thinspace^{\Sigma}\mathbf{\rho}-2D_{k}D^{k}\Omega+\Omega^{2}-\frac{1}{4}{}^{\Sigma}\mathbf{\rho}^{2}\right)+\frac{2}{\sqrt{\gamma}}\left(\left(\Pi_{ij}^{2}+\Pi^{2}\right)\left(\thinspace^{\Sigma}\mathbf{\rho}-2\Omega\right)-p^{kl}\Pi_{kl}\right),\label{aaaaaa}
\end{equation}
\end{tcolorbox}
\hphantom{}\\
and the momentum constraint (\ref{momentum=000020constraint}) becomes
\begin{tcolorbox}
\begin{equation}
\varPhi_{i}=-2\sqrt{\gamma}D_{k}\left(\frac{p_{i}^{k}}{\sqrt{\gamma}}\right)+2\Pi D_{i}\left(\thinspace^{\Sigma}\mathbf{\rho}-2\Omega\right)+4\sqrt{\gamma}D_{k}\left(\frac{\Pi_{i}^{k}}{\sqrt{\gamma}}\left(\thinspace^{\Sigma}\mathbf{\rho}-2\Omega\right)\right).\label{aaaaaa-1}
\end{equation}
\end{tcolorbox}
\hphantom{}

\subsection{The $R_{\mu\nu}R^{\mu\nu}$ theory }

In this section, we are going to evaluate the $R_{\mu\nu}R^{\mu\nu}$
theory as an example. To be able to do this, first, we have to compute
the space and time decomposition of the contraction $\rho_{\mu\nu}\rho^{\mu\nu}$.
We have 
\begin{equation}
f\left(\rho_{\mu\nu\rho\sigma}\right)=\rho_{\mu\nu}\rho^{\mu\nu}=\rho_{00}\rho^{00}+\rho_{0i}\rho^{0i}+\rho_{i0}\rho^{i0}+\rho_{ij}\rho^{ij},
\end{equation}
and 
\begin{equation}
f\left(\rho_{\mu\nu\rho\sigma}\right)=\rho_{00}\rho^{00}+2\rho_{0i}\rho^{0i}+\rho_{ij}\rho^{ij}.\label{f}
\end{equation}
Now we should decompose the corresponding components into the ADM
variables. We start with $\rho_{\mu\nu}$, which can be obtained as
follows 
\begin{equation}
\begin{aligned}\rho_{\mu\nu} & =g^{\alpha\beta}\rho_{\alpha\mu\beta\nu}\\
 & =g^{0\beta}\rho_{0\mu\beta\nu}+g^{i\beta}\rho_{i\mu\beta\nu}\\
 & =g^{00}\rho_{0\mu0\nu}+g^{0i}\rho_{0\mu i\nu}+g^{i0}\rho_{i\mu0\nu}+g^{ij}\rho_{i\mu j\nu}.
\end{aligned}
\end{equation}
Using the symmetries of $\rho_{\mu\nu g\sigma,}$ one has 
\begin{equation}
\rho_{\mu\nu}=g^{00}\rho_{0\mu0\nu}+g^{0i}\bigl(\rho_{\nu i\mu0}+\rho_{\mu i\nu0}\bigr)+g^{ij}\rho_{i\mu j\nu},
\end{equation}
which yields 
\begin{equation}
\rho_{00}=g^{ij}\rho_{i0j0}.\label{00}
\end{equation}
Recall that $\rho_{0000}$ and $\rho_{0i00}$ automatically vanish
because of the symmetries. Similarly $\rho_{\text{0i }}$ reads 
\begin{equation}
\rho_{0i}=-g^{0k}\rho_{i0k0}+g^{kl}\rho_{kil0}.\label{0i}
\end{equation}
Moreover, the spatial component $\rho_{ij}$ can be written as 
\begin{equation}
\rho_{ij}=g^{00}\rho_{0i0j}+g^{0k}\bigl(\rho_{ikj0}+\rho_{jki0}\bigr)+g^{kl}\rho_{kilj}.\label{ij}
\end{equation}
Recall that, we have already introduced the hypersurface projected
tensor fields $\rho_{ijk}$ and $\Omega_{ij}$ via (\ref{28})
and the inverse metric components. 
Let us reexpress $\rho_{00}$. Using (\ref{00}) we can write 
\begin{equation}
\rho_{00}=\Bigl(\gamma^{mn}-\frac{N^{m}N^{n}}{N^{2}}\Bigr)\Bigl(N^{2}\Omega_{mn}+N^{k}N^{l}\rho_{mknl}+NN^{k}\left(\rho_{mkn}+\rho_{nkm}\right)\Bigr),
\end{equation}
which yields

\begin{equation}
\begin{aligned}\rho_{00}= & N^{2}\gamma^{mn}\Omega_{mn}+N^{k}N^{l}\gamma^{mn}\rho_{mknl}+2NN^{k}\gamma^{mn}\rho_{mkn}-N^{m}N^{n}\Omega_{mn}\\
 & -\frac{N^{m}N^{n}}{N^{2}}N^{k}N^{l}\rho_{mknl}-\frac{N^{m}N^{n}}{N^ {}}N^{k}\rho_{mkn},
\end{aligned}
\end{equation}
where the last two terms vanish because of the symmetries. For simplicity,
we introduce 
\begin{eqnarray*}
\Omega\equiv\gamma^{mn}\Omega_{mn}, & ^{\Sigma}\rho_{kl}\equiv\gamma^{mn}\rho_{mknl}, & ^{\Sigma}\rho_{k}\equiv\gamma^{mn}\rho_{mkn},
\end{eqnarray*}
where $\rho_{mknl}$ and $\rho_{mkn}$ are purely spatial by assumption
and therefore we removed the over $\Sigma$ on these fields. Then,
$\rho_{00}$ reduces to 
\begin{equation}
\rho_{00}=N^{2}\Omega+N^{k}N^{l}\left(^{\Sigma}\rho_{kl}-\Omega_{kl}\right)+2NN^{k}\,{}^{\Sigma}\rho_{k}.\label{rho00}
\end{equation}
Now let us compute $\rho_{\text{0i }}$. One has (\ref{0i}), which
yields 
\begin{equation}
\begin{aligned}\rho_{oi}= & -\frac{N^{m}}{N^{2}}\Bigl(N^{2}\Omega_{mi}+N^{k}N^{l}\rho_{mkil}+NN^{k}\left(\rho_{mki}+\rho_{ikm}\right)\Bigr)\\
 & +\bigl(\gamma^{mn}-\frac{N^{m}N^{n}}{N^{2}}\bigr)\left(N\rho_{nim}+N^{l}\rho_{niml}\right),
\end{aligned}
\end{equation}
and
\begin{equation}
\begin{aligned}\rho_{0i}= & -N^{m}\Omega_{mi}-\frac{N^{m}N^{k}N^{l}}{N^{2}}\rho_{mkil}-\frac{N^{m}N^{k}}{N}\left(\rho_{mki}+\rho_{ikm}\right)\\
 & +N\rho_{ni}\ ^{n}+N^{l}\rho_{ni}\ ^{n}\ _{l}-\frac{N^{m}N^{n}\rho_{nim}}{N}N_{n}-\frac{N^{m}N^{n}N^{l}\rho_{niml}}{N^{2}}.
\end{aligned}
\end{equation}
Then it reduces to 
\begin{equation}
\rho_{0i}=-N^{m}\Omega_{mi}-\frac{N^{m}N^{k}\rho_{ikm}}{N^{2}}+N\ {}^{\Sigma}\rho_{i}+N^{m}{}^{\Sigma}\rho_{im}-\frac{N^{m}N^{k}\rho_{kim}}{N},
\end{equation}
where $\rho_{ikm}=-\rho_{\text{kim }}$. Then, one ends up with 
\begin{equation}
\rho_{0i}=N\ {}^{\Sigma}\rho_{i}+N^{k}\bigl({}^{\Sigma}\rho_{ik}-\Omega_{ik}\bigr).\label{rho0i}
\end{equation}
Similarly we can compute $\rho_{ij}$. Using (\ref{ij}), one has
\begin{equation}
\begin{aligned}\rho_{ij}= & -\frac{1}{N^{2}}\left(N^{2}\Omega_{ij}+N^{k}N^{l}\rho_{ikjl}+NN^{k}\left(\rho_{ikj}+\rho_{jki}\right)\right)\\
 & +\frac{N^{m}}{N^{2}}\left(N\bigl(\rho_{jmi}+\rho_{imj}\bigr)+N^{l}\bigl(\rho_{jmil}+\rho_{imjl}\bigr)\right)+\Bigl(\gamma^{mn}-\frac{N^{m}N^{n}}{N^{2}}\Bigr)\rho_{minj}.
\end{aligned}
\end{equation}
Then, we obtain 
\begin{equation}
\begin{aligned}\rho_{ij}= & -\Omega_{ij}-\frac{N^{k}N^{l}}{N^{2}}\rho_{ikjl}-\frac{N^{k}}{N}\left(\rho_{ikj}+\rho_{jki}\right)+\frac{N^{m}}{N}\left(\rho_{jmi}+\rho_{imj}\right)\\
 & +\frac{N^{k}N^{l}}{N^{2}}\left(\rho_{jkil}+\rho_{ikjl}\right)+^ {}\rho_{mi}{}^{m}\ _{j}-\frac{N^{k}N^{l}}{N^{2}}\rho_{kilj},
\end{aligned}
\end{equation}
which reduces to the following compact form 
\begin{equation}
\rho_{ij}={}^{\Sigma}\rho_{ij}-\Omega_{ij}.\label{rhoij}
\end{equation}
To compute the contraction $\rho_{\mu\nu}\rho^{\mu\nu}$, we have
to compute the higher indices versions of the components. Let us start
with $\rho^{00}$ . One has 
\begin{equation}
\begin{aligned}\rho^{00}= & g^{\mu0}g^{0\nu}\rho_{\mu\nu}=g^{00}g^{00}\rho_{00}+2g^{m0}g^{00}\rho_{m0}+g^{n0}g^{m0}\rho_{nm}.\end{aligned}
\end{equation}
Inserting the inverse spacetime metric components, we have 
\begin{equation}
\rho^{00}=\frac{1}{N^{4}}\rho_{00}-\frac{2N^{m}}{N^{4}}\rho_{0m}+\frac{N^{n}N^{m}\rho_{nm}}{N^{4}},
\end{equation}
and making use of (\ref{rho00}, \ref{rho0i}, \ref{rhoij}) one arrives
at 
\begin{equation}
\begin{aligned} & \rho^{00}=\frac{1}{N^{4}}\biggl(N^{2}\Omega+N^{k}N^{l}\left(^{\Sigma}\rho_{kl}-\Omega_{kl}\right)+2NN^{k}\ ^{\Sigma}\rho_{k}\\
 & -2NN^{m}\ ^{\Sigma}\rho_{m}-2N^{m}N^{k}\left(^{\Sigma}\rho_{km}-\Omega_{km}\right)+N^{m}N^{n}\left(^{\Sigma}\rho_{nm}-\Omega_{mn}\right)\biggr),
\end{aligned}
\end{equation}
where all the terms, except the first one on the right-hand side of
the last equation cancels each other. Therefore we arrive at a simple
result
\begin{equation}
\rho^{00}=\frac{\Omega}{N^{2}}.
\end{equation}
Now we can compute the first piece in (\ref{f}), that is $\rho_{00}\rho^{00}$.
One has the following 
\begin{equation}
\rho_{00}\rho^{00}=\Omega^{2}+\frac{N^{k}N^{l}}{N}\Omega\left(^{\Sigma}\rho_{kl}-\Omega_{kl}\right)+2\frac{N^{k}}{N}\Omega\ {}^{\Sigma}\rho_{k}.\label{fisrt=00003D000020piece}
\end{equation}
Similarly, we can evaluate $\rho^{0i}$:
\begin{equation}
\begin{aligned}\rho^{0i}=g^{\mu0}g^{\nu i}\rho_{\mu\nu}=g^{00}g^{0i}\rho_{00}+g^{m0}g^{0i}\rho_{m0}+g^{00}g^{mi}\rho_{0m}+g^{n0}g^{mi}\rho_{nm},\end{aligned}
\end{equation}
and we can write 
\begin{equation}
\rho^{0i}=g^{00}g^{0i}\rho_{00}+\rho_{m0}\left(g^{00}g^{mi}+g^{m0}g^{0i}\right)+g^{n0}g^{mi}\rho_{nm}.
\end{equation}
More explicitly, one obtains
\begin{equation}
\begin{aligned}\rho^{0i}= & g^{00}g^{0i}N^{2}\Omega\\
 & +\left(^{\Sigma}\rho_{kl}-\Omega_{kl}\right)\Bigl(N^{k}N^{l}g^{00}g^{0i}+g^{k0}g^{li}+N^{k}\left(g^{00}g^{li}+g^{l0}g^{0i}\right)\Bigr)\\
 & +\ ^{\Sigma}\rho_{k}\left(2NN^{k}g^{00}g^{0i}+N\left(g^{00}g^{ki}+g^{k0}g^{0i}\right)\right).
\end{aligned}
\end{equation}
Inserting the inverse metric components, one ends up with 
\begin{equation}
\rho^{0i}=-\frac{1}{N}\bigl(N^{i}\ \Omega+\ ^{\Sigma}\rho^{i}\bigr).
\end{equation}
Then, the second piece in (\ref{f}) becomes 
\begin{eqnarray}
\rho_{0i}\rho^{0i} & = & -^{\Sigma}\rho_{i}\ ^{\Sigma}\rho^{i}-\frac{1}{N}N^{i}\ ^{\Sigma}\rho_{i}\ \Omega-\frac{N^{i}N^{k}}{N^{2}}\Omega\bigl(\ ^{\Sigma}\rho_{ik}-\Omega_{ik}\bigr)\nonumber \\
 &  & -\frac{N^{k}}{N}\ ^{\Sigma}\rho^{i}\left(^{\Sigma}\ \rho_{ik}-\Omega_{ik}\right).
\end{eqnarray}
Note that $\rho_{ij}\neq\ ^{\Sigma}\rho_{ij}$. Let's 
continue with $\rho^{{ij}}$. One can express 
\begin{equation}
\begin{aligned}\rho^{ij} & =g^{i\mu}g^{j\nu}\rho_{\mu\nu}\\
 & =g^{i0}g^{j0}\rho_{00}+\rho_{m0}\bigl(g^{im}g^{j0}+g^{i0}g^{jm}\bigr)+g^{in}g^{jm}\rho_{nm}.
\end{aligned}
\end{equation}
Since we will compute the contraction $\rho_{ij}\rho^{ij}$, we may use the symmetries of the indices at this step to simplify the construction from now on. Inserting the results (\ref{rho00}, \ref{rho0i}, \ref{rhoij}),
we obtain
\begin{equation}
\begin{aligned}\rho^{ij}= & N^{2}g^{i0}g^{j0}\ \Omega+2N\ ^{\Sigma}\rho_{k}\left(N^{k}g^{i0}g^{j0}+g^{jk}g^{i0}\right)\\
 & +\left(^{\Sigma}\rho_{kl}-\Omega_{kl}\right)\left(g^{i0}g^{j0}N^{k}N^{l}+2N^{k}g^{i0}g^{jl}+g^{ik}g^{jl}\right).
\end{aligned}
\end{equation}
After using the inverse metric components, the last equation reduces
to 
\begin{equation}
\rho^{ij}=\frac{N^{i}N^{j}}{N^{2}}\Omega+\ ^{\Sigma}\rho^{ij}-\Omega^{ij}+\frac{2}{N}N^{i}\ ^{\Sigma}\rho^{j}.
\end{equation}
The last term in (\ref{f}) is easy to construct. We can easily obtain
\begin{align*}
\rho_{ij}\rho^{ij}=\left(\ ^{\Sigma}\rho_{ij}-\Omega_{ij}\right)^{2} & +\frac{N^{i}N^{j}}{N^{2}}\Omega\left(\ ^{\Sigma}\rho_{ij}-\Omega_{ij}\right)+\frac{2}{N}N^{i}\ ^{\Sigma}\rho^{j}\left(\ ^{\Sigma}\rho_{ij}-\Omega_{ij}\right),
\end{align*}
where 
\begin{equation}
\left(\ ^{\Sigma}\rho_{ij}-\Omega_{ij}\right)^{2}=\bigl(\ ^{\Sigma}\rho_{ij}-\Omega_{ij}\bigr)\bigl(\ ^{\Sigma}\rho^{ij}-\Omega^{ij}\bigr).
\end{equation}
Collecting the pieces, $\rho_{\mu\nu}\rho^{\mu\nu}$ becomes

\begin{tcolorbox}
\begin{equation}
f\left(\rho_{\mu\nu\rho\sigma}\right)=\rho_{\mu\nu}\rho^{\mu\nu}=\Omega^{2}-2^{\Sigma}\rho_{i}\ ^{\Sigma}\rho^{i}+\ ^{\Sigma}\rho_{ij}\ ^{\Sigma}\rho^{ij}-2\Omega^{ij}\ ^{\Sigma}\rho_{ij}+\Omega_{ij}\ \Omega^{ij}.\label{f_explicit}
\end{equation}
\end{tcolorbox}
\hphantom{}\\
Recall that the constraint on the auxiliary field $\psi$ was given
in (\ref{constraint=000020on=000020psi}). In our case, differentiation
of $f$ with respect to $\Omega_{ij}$ yields 
\begin{eqnarray}
\frac{\partial f}{\partial\Omega_{ij}} & = & \frac{\partial}{\partial\Omega_{ij}}\biggl(\gamma^{kl}\gamma^{mn}\Omega_{k1}\Omega_{mn}-2^{\Sigma}\rho_{i}\ ^{\Sigma}\rho^{i}+\ ^{\Sigma}\rho_{ij}\ ^{\Sigma}\rho^{ij}\nonumber \\
 &  & -2^{\Sigma}\rho^{mn}\Omega_{mn}+\Omega_{mn}\Omega_{kl}\gamma^{km}\gamma^{ln}\biggr).
\end{eqnarray}
Working out the details, one has
\begin{eqnarray}
\frac{\partial f}{\partial\Omega_{ij}} & = & \gamma^{kl}\gamma^{mn}\left(\Omega_{kl}\frac{\partial\Omega_{mn}}{\partial\Omega_{ij}}+\Omega_{mn}\frac{\partial\Omega_{k1}}{\partial\Omega_{ij}}\right)-2\ ^{\Sigma}\rho^{mn}\frac{\partial\Omega_{mn}}{\partial\Omega_{ij}}\nonumber \\
 &  & +\gamma^{km}\gamma^{ln}\left(\frac{\partial\Omega_{mn}}{\partial\Omega_{ij}}\Omega_{kl}+\frac{\partial\Omega_{kl}}{\partial\Omega_{ij}}\Omega_{mn}\right),
\end{eqnarray}
and using 
\begin{equation}
\frac{\partial\Omega_{mn}}{\partial\Omega_{ij}}=\frac{1}{2}\left(\delta_{m}^{i}\delta_{n}^{j}+\delta_{n}^{i}\delta_{m}^{j}\right),
\end{equation}
one arrives at 
\begin{equation}
\frac{\partial f}{\partial\Omega_{ij}}=2\left(\Omega^{ij}+\Omega\gamma^{ij}-\ ^{\Sigma}\rho^{ij}\right).
\end{equation}
Then, the constraint equation of the auxiliary field (\ref{constraint=000020on=000020psi})
yields
\begin{tcolorbox}
\begin{equation}
\psi^{ij}=\ ^{\Sigma}\rho^{ij}-\Omega\gamma^{ij}-\Omega^{ij}.
\end{equation}
\end{tcolorbox}
\hphantom{}\\
We have already introduced the Hamiltonian and the momentum constraint
equations of generic $f$(Riemann) theories in (\ref{Hamiltonian=000020constraint},
\ref{momentum=000020constraint}). Using these expressions, the momentum
constraint reduces to
\begin{tcolorbox}
\begin{align}
\Phi_{i}= & -2\sqrt{\gamma}D_{k}\left(\frac{P_{i}{}^{k}}{\sqrt{\gamma}}\right)+\Pi_{kl}D_{i}\left(^{\Sigma}\rho^{kl}-\Omega^{kl}\right)-\Pi D_{i}\Omega\\
 & +2\sqrt{\gamma}D_{k}\left(\frac{\Pi_{li}}{\sqrt{\gamma}}\left(^{\Sigma}\rho^{kl}-\Omega^{kl}\right)\right)-2\sqrt{\gamma}D^{k}\left(\frac{\Omega\Pi_{ki}}{\sqrt{\gamma}}\right).\nonumber 
\end{align}
\end{tcolorbox}
\vphantom{} \\
Similarly, the Hamiltonian constraint becomes 

\begin{tcolorbox}
\begin{eqnarray}
\Phi_{0} & = & {\displaystyle \frac{\sqrt{\gamma}}{2}}\Bigl(\Omega^{2}+2{}^{\Sigma}\rho_{i}\ ^{\Sigma}\rho^{i}-\ ^{\Sigma}\rho_{ij}^{2}+\Omega_{ij}^{2}+2D_{i}D_{j}\left(^{\Sigma}\rho^{ij}-\Omega^{ij}\right)-2D_{i}D^{i}\Omega\Bigr)\nonumber \\
 &  & +\frac{1}{\sqrt{\gamma}}\Bigl(-2\Pi_{ij}p^{ij}+\left(^{\Sigma}\rho^{ij}-\Omega^{ij}\right)\left(\Pi_{ij}+\Pi_{ik}\Pi_{j}^{k}\right)-\Omega\left(\Pi^{2}+\Pi_{ij}^{2}\right)\Bigr).
\end{eqnarray}
\end{tcolorbox}

\section{Conclusions}

We studied the time evolution and the constraint structure of $f(Riemann)$-type theories using the auxiliary fields as was done in \cite{Deruelle}
and recast the Hamiltonian flow in the compact Fischer-Marsden form
\cite{Fischer-Marsden}. This form of Einstein's equations can be considered to be a failure of the initial data to possess an exact time translation symmetry, a vantage point that led to a definition of non-stationary energy or Dain's invariant \cite{Dain}. The type of theories we studied here represent a large class of theories that can be handled with two auxiliary fields, going beyond these and including the derivatives of the Riemann tensor is somewhat challenging which we shall do in a separate work. One of our motivations for this work was to give a detailed account of the computations leading to the final constraints and time evolution expressions as there are several important mistakes and omissions in the existing literature. We also gave two concrete examples: the $R^2$ and $R_{\mu \nu} R^{\mu \nu}$ theories.

\acknowledgments

E. Altas is supported by the TUBITAK Grant No. 123F353.

\section{Appendices}

\appendix

\section{ADM decomposition}

\subsection{The metric and the connection}

For the sake of completeness let us give here the ADM split of the
Einstein's equations and all the relevant tensors. Using the $\left(n-1\right)+1$
dimensional splitting of the spacetime metric (\ref{ADMdecompositionofmetric}),
we can express 
\begin{equation}
g_{00}=-N^{2}+N_{i}N^{i},~\ ~~~g_{0i}=N_{i},~~~~\ g_{ij}=\gamma_{ij},
\end{equation}
and the inverse metric components are 
\begin{equation}
g^{00}=-\frac{1}{N^{2}},~~~~g^{0i}=\frac{1}{N^{2}}N^{i},~~~~g^{ij}=\gamma^{ij}-\frac{1}{N^{2}}N^{i}N^{j}.
\end{equation}
In generic $n$ dimensions, the spacetime metric in a matrix form
reads 
\begin{equation}
g_{\mu\nu}=\begin{pmatrix}g_{00} & g_{0i}\\
g_{0i} & g_{ij}
\end{pmatrix}=\begin{pmatrix}N_{i}N_{i}-N^{2} & N_{i}\\
N_{i} & \gamma_{ij}
\end{pmatrix}.
\end{equation}
Taking the determinant of the spacetime metric, we can relate the
determinants of the spacetime metric and the spatial metric as 
\begin{equation}
\sqrt{-g}=N\sqrt{\gamma},
\end{equation}
where we have used $g=\det g_{\mu\nu}$ and similarly $\gamma=\det\gamma_{ij}$.
Let $\Gamma_{\nu\rho}^{\mu}$ denote the Christoffel symbol of the
$n$ dimensional spacetime 
\begin{equation}
\Gamma_{\nu\rho}^{\mu}=\frac{1}{2}g^{\mu\sigma}\left(\partial_{\nu}g_{\rho\sigma}+\partial_{\rho}g_{\nu\sigma}-\partial_{\sigma}g_{\nu\rho}\right),
\end{equation}
and let $^{\Sigma}\Gamma_{ij}^{k}$ denote the Christoffel symbol
of the $n-1$ dimensional hypersurface, which is compatible with the
spatial metric $\gamma_{ij}$, $D_{k}\gamma_{ij}=0$, as 
\begin{equation}
^{\Sigma}\Gamma_{ij}^{k}=\frac{1}{2}\gamma^{kl}\left(\partial_{i}\gamma_{jl}+\partial_{j}\gamma_{il}-\partial_{l}\gamma_{ij}\right).\label{Christoffelofhypersurface}
\end{equation}
Then a simple computation gives one the following components of the
connection 
\begin{equation}
\Gamma_{00}^{0}=\frac{1}{N}\left(\dot{N}+N^{k}(\partial_{k}N+N^{i}K_{ik})\right),
\end{equation}
and 
\begin{equation}
\Gamma_{0i}^{0}=\frac{1}{N}\left(\partial_{i}N+N^{k}K_{ik}\right),~~~~\Gamma_{ij}^{0}=\frac{1}{N}K_{ij},~~~~\Gamma_{ij}^{k}=\thinspace^{\Sigma}\Gamma_{ij}^{k}-\frac{N^{k}}{N}K_{ij},
\end{equation}
and 
\begin{equation}
\Gamma_{0j}^{i}=-\frac{1}{N}N^{i}\left(\partial_{j}N+K_{kj}N^{k}\right)+NK_{j}\thinspace^{i}+D_{j}N^{i},
\end{equation}
and also 
\begin{equation}
\Gamma_{00}^{i}=-\frac{N^{i}}{N}\left(\dot{N}+N^{k}\left(\partial_{k}N+N^{l}K_{kl}\right)\right)+N\left(\partial^{i}N+2N^{k}K_{k}\thinspace^{i}\right)+\dot{N}^{i}+N^{k}D_{k}N^{i}.
\end{equation}
To compute the decomposition of the field equations, we need to express
the additional tensor quantities such as the Riemann and the Ricci
tensor components, the scalar curvature.

\subsection{{\normalsize{}{}{}{}{}{}{}ADM splitting of the Riemann tensor}}

The Riemann tensor of the spacetime is defined as 
\begin{equation}
R^{\mu}\thinspace_{\nu\rho\sigma}=\partial_{\rho}\Gamma_{\nu\sigma}^{\mu}-\partial_{\sigma}\Gamma_{\nu\rho}^{\mu}+\Gamma_{\rho\gamma}^{\mu}\Gamma_{\nu\sigma}^{\gamma}-\Gamma_{\sigma\gamma}^{\mu}\Gamma_{\nu\rho}^{\gamma}.
\end{equation}
So, it is straightforward to compute the components given below 
\begin{equation}
R^{m}\thinspace_{jkl}={}^{\Sigma}R^{m}\thinspace_{jkl}+K_{jl}K_{k}^{m}-K_{jk}K_{l}^{m}+\frac{N^{m}}{N}\left(D_{l}K_{jk}-D_{k}K_{jl}\right),
\end{equation}
\begin{equation}
R^{0}\thinspace_{jkl}=\frac{1}{N}\left(D_{k}K_{jl}-D_{j}K_{kl}\right),
\end{equation}
\begin{equation}
R^{0}\thinspace_{j0i}=\frac{1}{N}\left(\dot{K}_{ij}-D_{i}D_{j}N-N^{l}D_{i}K_{jl}-2K_{l(i}D_{j)}N^{l}\right)-K_{jk}K_{i}^{k},
\end{equation}
where $^{\Sigma}R^{m}\thinspace_{jkl}$ is the Riemann tensor of the
hypersurface and it explicitly reads 
\begin{equation}
^{\Sigma}R^{m}\thinspace_{jkl}=\partial_{k}\thinspace^{\Sigma}\Gamma_{jl}^{m}-\partial_{l}\thinspace^{\Sigma}\Gamma_{jk}^{m}+\thinspace^{\Sigma}\Gamma_{ks}^{m}\thinspace^{\Sigma}\Gamma_{jl}^{s}-\thinspace^{\Sigma}\Gamma_{ls}^{m}\thinspace^{\Sigma}\Gamma_{jk}^{s}.
\end{equation}
Also, we need to compute $R_{i0j0}$. Using the above results it becomes
\begin{align}
R_{i0j0} & =N^{k}N^{l}\thinspace{}^{\Sigma}R\thinspace_{ikjl}+NN^{k}\left(D_{i}K_{jk}+D_{j}K_{ik}-2D_{k}K_{ij}\right)\nonumber \\
 & ~~~~~~~~~~+N\left(D_{i}D_{j}N-\dot{K}_{ij}+\mathcal{L_{N}}K_{ij}\right)+N^{2}K_{ik}K_{j}^{k},
\end{align}
where $\mathcal{L_{N}}$ denotes the Lie differentiation along the
shift vector $N^{i}$ and when it acts on the extrinsic curvature,
one has 
\begin{equation}
\mathcal{L_{N}}K_{ij}=N^{k}D_{k}K_{ij}+K_{ki}D_{j}N^{k}+K_{kj}D_{i}N^{k}.
\end{equation}
Moreover, we can introduce the hypersurface projected components of
the Riemann tensor as $R_{ijkl}$, $R_{ijk\vec{n}}$ and $R_{i\vec{n}j\vec{n}}$.
Below we will prove the following three statements: 
\begin{equation}
R_{ijkl}={}^{\Sigma}R_{ijkl}+K_{ik}K_{jl}-K_{il}K_{jk},\label{birinci}
\end{equation}
\begin{equation}
R_{ijk\vec{n}}=n^{\mu}R_{ijk\mu}=D_{i}K_{jk}-D_{j}K_{ik},\label{ikinci}
\end{equation}
\begin{equation}
R_{i\vec{n}j\vec{n}}=n^{\mu}n^{\nu}R_{i\mu j\nu}=\frac{1}{N}\left(\mathcal{L_{N}}K_{ij}+D_{i}D_{j}N-\dot{K}_{ij}\right)+K_{ik}K_{j}^{k}.\label{ucuncu}
\end{equation}
Let us start with the proof of the first statement (\ref{birinci}).
We have

\begin{equation}
R_{ijkl}=g_{i\mu}R^{\mu}{}_{jkl}=g_{i0}R^{0}\thinspace_{jkl}+g_{im}R^{m}{}_{jkl}=N_{i}R^{0}\thinspace_{jkl}+\gamma_{im}R^{m}{}_{jkl},
\end{equation}
where one can express 
\begin{eqnarray}
R^{0}\,_{jkl} & = & \partial_{k}\Gamma_{jl}^{0}-\partial_{l}\Gamma_{jk}^{0}+\Gamma_{k\mu}^{0}\Gamma_{jl}^{\mu}-\Gamma_{l\mu}^{0}\Gamma_{jk}^{\mu}\nonumber \\
 & = & \partial_{k}\left(\frac{1}{N}K_{jl}\right)-\partial_{l}\left(\frac{1}{N}K_{jk}\right)+\frac{1}{N^{2}}K_{\dot{j}l}\left(\partial_{k}N+N^{m}K_{mk}\right)\nonumber \\
 &  & +\frac{1}{N}K_{km}\left(^{\Sigma}\Gamma_{jl}^{m}-\frac{N^{m}}{N}K_{\dot{j}l}\right)-\frac{1}{N^{2}}K_{\dot{j}k}\left(\partial_{l}N+N^{m}K_{ml}\right)\nonumber \\
 &  & -\frac{1}{N}K_{lm}\left(^{\Sigma}\Gamma_{\dot{j}k}^{m}-\frac{N^{m}}{N}K_{jk}\right),
\end{eqnarray}
which yields

\begin{equation}
R^{0}\,_{jkl}=\frac{1}{N}\left(\partial_{k}K_{jl}-\partial_{l}K_{jk}+K_{km}{}^{\Sigma}\Gamma_{jl}^{m}-K_{lm}{}^{\Sigma}\Gamma_{jk}^{m}\right),
\end{equation}
or in terms of the hypersurface covariant derivatives

\begin{equation}
R^{0}{}_{jkl}=\frac{1}{N}\left(D_{k}K_{jl}-D_{l}K_{jk}\right).
\end{equation}
Similarly we can compute $R^{m}{}_{jkl}$. By definition we have

\begin{equation}
R^{m}{}_{jkl}=\partial_{k}\Gamma_{\dot{j}l}^{m}-\partial_{l}\Gamma_{\dot{j}k}^{m}+\Gamma_{k\mu}^{m}\Gamma_{jl}^{\mu}-\Gamma_{l\mu}^{m}\Gamma_{jk}^{\mu},
\end{equation}
which reduces to 
\begin{equation}
R^{m}{}_{jkl}={}^{\Sigma}R^{m}{}_{jkl}+K_{jl}K_{k}{}^{m}-K_{jk}K_{l}^{m}+\frac{N^{m}}{N}\left(D_{l}K_{jk}-D_{k}K_{jl}\right).
\end{equation}
Collecting the pieces we end up with (\ref{birinci}).

Let us prove (\ref{ikinci}). By definition, projection once can be
written as

\begin{equation}
R_{ijk\vec{n}}=n^{\mu}R_{ijk\mu}=n^{0}R_{ijk0}+n^{l}R_{ijkl}=\frac{1}{N}R_{ijk0}-\frac{N^{l}}{N}R_{ijkl},\label{ara29}
\end{equation}
where 
\begin{eqnarray}
R_{ijk0}=g_{00}R^{0}\,_{kji}+g_{0l}R^{l}\,_{kji},
\end{eqnarray}
which reads 
\begin{equation}
R_{ijk0}=N^{l}\,{}^{\Sigma}R_{lkji}+N\left(D_{i}K_{jk}-D_{j}K_{ik}\right)+N^{l}\left(K_{ki}K_{jl}-K_{kj}K_{il}\right).
\end{equation}
Using the last equation in (\ref{ara29}), one arrives at the desired
result (\ref{ikinci}).

We also need to construct $R_{i\overrightarrow{n}j\overrightarrow{n}}$,
which reads

\begin{equation}
R_{i\vec{n}j\vec{n}}=n^{\mu}n^{\nu}R_{i\mu j\nu}=-R^{0}\,_{j0i}+N^{k}R^{0}\,_{jki},
\end{equation}
where the nonvanishing components of the first term are 
\begin{equation}
R^{0}\,_{j0i}=\frac{1}{N}\left[\dot{K}_{ij}-D_{i}D_{j}N-N^{l}D_{i}K_{lj}-K_{lj}D_{i}N^{l}-K_{ik}D_{j}N^{k}\right]-K_{ik}K_{j}^{k}.
\end{equation}
Substituting the results, we obtain the desired result (\ref{ucuncu}).

\subsection{{\normalsize{}{}{}{}{}{}{}ADM splitting of the Ricci tensor
and the scalar curvature}}

Starting with the definition of the spacetime Ricci tensor 
\begin{equation}
R_{\rho\sigma}=\partial_{\mu}\Gamma_{\rho\sigma}^{\mu}-\partial_{\rho}\Gamma_{\mu\sigma}^{\mu}+\Gamma_{\mu\nu}^{\mu}\Gamma_{\rho\sigma}^{\nu}-\Gamma_{\sigma\nu}^{\mu}\Gamma_{\mu\rho}^{\nu},
\end{equation}
one can express 
\begin{eqnarray*}
 &  & R_{ij}=\partial_{0}\Gamma_{ij}^{0}+\partial_{k}\Gamma_{ij}^{k}-\partial_{i}(\Gamma_{0j}^{0}+\Gamma_{kj}^{k})+\Gamma_{ij}^{0}(\Gamma_{00}^{0}+\Gamma_{k0}^{k})\\
 &  & ~~~~~~+\Gamma_{ij}^{k}\Gamma_{0k}^{0}+\Gamma_{kl}^{k}\Gamma_{ij}^{l}-\Gamma_{0j}^{0}\Gamma_{0i}^{0}-\Gamma_{kj}^{0}\Gamma_{0i}^{k}-\Gamma_{ki}^{0}\Gamma_{0j}^{k}-\Gamma_{jl}^{k}\Gamma_{ki}^{l},
\end{eqnarray*}
which yields

\begin{equation}
R_{ij}={}^{\Sigma}R_{ij}+KK_{ij}-2K_{ik}K_{j}^{k}+\frac{1}{N}\left(\dot{K}_{ij}-N^{k}D_{k}K_{ij}-D_{i}D_{j}N-K_{ki}D_{j}N^{k}-K_{kj}D_{i}N^{k}\right),\label{eq:rij}
\end{equation}
where $^{\Sigma}R_{ij}$ denotes the $ij$ component of the Ricci
tensor on the hypersurface 
\begin{equation}
^{\Sigma}R_{ij}=\partial_{k}\thinspace^{\Sigma}\Gamma_{ij}^{k}-\partial_{i}\thinspace^{\Sigma}\Gamma_{kj}^{k}+\thinspace^{\Sigma}\Gamma_{kl}^{k}\thinspace^{\Sigma}\Gamma_{ij}^{l}-\thinspace^{\Sigma}\Gamma_{jl}^{k}\thinspace^{\Sigma}\Gamma_{ki}^{l}.
\end{equation}
The $0i$ component can be written as 
\begin{equation}
R_{0i}=\partial_{0}\Gamma_{0i}^{0}+\partial_{k}\Gamma_{0i}^{k}-\partial_{i}(\Gamma_{00}^{0}+\Gamma_{k0}^{k})+\Gamma_{0i}^{0}\Gamma_{k0}^{k}+\Gamma_{kl}^{k}\Gamma_{i0}^{l}-\Gamma_{00}^{k}\Gamma_{ki}^{0}-\Gamma_{0l}^{k}\Gamma_{ki}^{l},
\end{equation}
and this expression gives us the following simple result 
\begin{equation}
R_{0i}=N^{j}R_{ij}+N\left(D_{m}K_{i}^{m}-D_{i}K\right).\label{eq:ri0}
\end{equation}
Similarly, the $00$ component 
\begin{equation}
R_{00}=\partial_{k}\Gamma_{00}^{k}-\partial_{0}\Gamma_{0k}^{k}+\Gamma_{00}^{0}\Gamma_{k0}^{k}+\Gamma_{kl}^{k}\Gamma_{00}^{l}-\Gamma_{00}^{k}\Gamma_{k0}^{0}-\Gamma_{0l}^{k}\Gamma_{k0}^{l},
\end{equation}
can be written in a compact form as

\begin{equation}
R_{00}=N^{i}N^{j}R_{ij}-N^{2}K_{ij}K^{ij}+N\left(D_{k}D^{k}N-\dot{K}-N^{k}D_{k}K+2N^{k}D_{m}K_{k}^{m}\right).\label{r00}
\end{equation}
Then, the scalar curvature of the spacetime, $R=g^{\mu\nu}R_{\mu\nu}$,
can be expressed in terms of the scalar curvature of the spatial hypersurface,
$\thinspace^{\Sigma}R=\gamma^{ij}~{}^{\Sigma}R_{ij}$, as

\begin{equation}
R=\thinspace^{\Sigma}R+K^{2}+K_{ij}K^{ij}+\frac{2}{N}\left(\dot{K}-D_{k}D^{k}N-N^{k}D_{k}K\right).\label{r}
\end{equation}

\section{Field equations of the $f$(Riemann) theory}

Let us start with the action 
\begin{equation}
S\left[g_{\mu v}\right]=\frac{1}{2}\int_{\mathcal{M}}d\thinspace^{n}x\sqrt{-g}~f\left(R_{\mu v\rho\sigma}\right),\label{denk1}
\end{equation}
of which the first-order variation is 
\begin{equation}
\delta S\left[g_{\mu v}\right]=\frac{1}{2}\int_{\mathcal{M}}d\thinspace^{n}x\left(\delta\sqrt{-g}~f\left(R_{\mu v\rho\sigma}\right)+\sqrt{-g}~\delta f\left(R_{\mu v\rho\sigma}\right)\right),\label{denk2}
\end{equation}
where 
\begin{equation}
\delta\sqrt{-g}=\frac{1}{2}\sqrt{-g}g^{\mu\nu}\delta g_{\mu\nu}=-\frac{1}{2}\sqrt{-g}g_{\mu\nu}\delta g^{\mu\nu},\label{variation=000020det}
\end{equation}
and 
\begin{equation}
\delta f\left(R_{\mu v\rho\sigma}\right)=\frac{\partial f}{\partial R_{\lambda\gamma\rho\sigma}}\delta\left(g_{\lambda\tau}R^{\tau}{}_{\gamma\rho\sigma}\right)=\frac{\partial f}{\partial R_{\lambda\gamma\rho\sigma}}\left(R^{\tau}\,_{\gamma\rho\sigma}\delta g_{\lambda\tau}+g_{\lambda\tau}\delta R^{\tau}\,_{\gamma\rho\sigma}\right).\label{denk4}
\end{equation}
Here, the linear order variation of the Riemann tensor is $\delta R^{\tau}\,_{\gamma\rho\sigma}=\nabla_{\rho}\delta\Gamma_{\gamma\sigma}^{\tau}-\nabla_{\sigma}\delta\Gamma_{\gamma\rho}^{\tau}$
and the variation of the spacetime connection is 
\begin{equation}
\delta\Gamma_{\mu v}^{\sigma}=\frac{1}{2}g^{\sigma\lambda}\left(\nabla_{\mu}\delta g_{v\lambda}+\nabla_{v}\delta g_{\mu\lambda}-\nabla_{\lambda}\delta g_{\mu v}\right).
\end{equation}
So we have 
\begin{eqnarray}
 & \delta f\left(R_{\mu\nu\rho\sigma}\right) & =\frac{\partial f}{\partial R_{\lambda\gamma\rho\sigma}}\left(R^{\tau}\,_{\gamma\rho\sigma}\delta g_{\lambda\tau}+g_{\lambda\tau}\left(\nabla_{\rho}\delta\Gamma_{\gamma\sigma}^{\tau}-\nabla_{\sigma}\delta\Gamma_{\gamma\rho}^{\tau}\right)\right)\nonumber \\
 &  & =\frac{\partial f}{\partial R_{\lambda\gamma\rho\sigma}}\left(R^{\tau}\,_{\gamma\rho\sigma}\delta g_{\lambda\tau}+\nabla_{\rho}\nabla_{[\gamma}\delta g_{\lambda]\sigma}+\nabla_{\sigma}\nabla_{[\lambda}\delta g_{\gamma]\rho}\right).
\end{eqnarray}
By renaming the indices and using the antisymmetry of the Riemann
tensor, the last equation can be written as 
\begin{equation}
\delta f\left(R_{\mu\nu\rho\sigma}\right)=\frac{\partial f}{\partial R_{\lambda\gamma\rho\sigma}}\left(R^{\tau}\,_{\gamma\rho\sigma}\delta g_{\lambda\tau}+2\nabla_{\rho}\nabla_{\gamma}\delta g_{\sigma\lambda}\right).
\end{equation}
Collecting the pieces, we arrive at 
\begin{equation}
\delta S\left[g_{\mu\nu}\right]=\frac{1}{2}\int_{\mathcal{M}}d\thinspace^{n}x\sqrt{-g}\biggl(\frac{1}{2}g^{\mu\nu}\delta g_{\mu\nu}~f\left(R_{\lambda\gamma\rho\sigma}\right)+\frac{\partial f}{\partial R_{\lambda\gamma\rho\sigma}}\left(R^{\tau}\,_{\gamma\rho\sigma}\delta g_{\lambda\tau}+2\nabla_{\rho}\nabla_{\gamma}\delta g_{\sigma\lambda}\right)\biggr).
\end{equation}
Using integration by parts and ignoring the boundary terms, one has
\begin{eqnarray}
\delta S\left[g_{\mu\nu}\right] & = & \frac{1}{4}\int_{\mathcal{M}}d\thinspace^{n}x\sqrt{-g}\Bigg(g^{\mu\nu}\delta g_{\mu\nu}~f{\left(R_{\lambda\gamma\rho\sigma}\right)}\\
 &  & \quad+\delta g_{\mu\nu}\left(\frac{\partial f}{\partial R_{\mu\gamma\rho\sigma}}R^{\nu}\,_{\gamma\rho\sigma}+\frac{\partial f}{\partial R_{\nu\gamma\rho\sigma}}R^{\mu}\,_{\gamma\rho\sigma}\right)+2\delta g_{\mu\nu}\nabla_{\sigma}\nabla_{\rho}\left(\frac{\partial f}{\partial R_{\mu\sigma\rho\nu}}+\frac{\partial f}{\partial R_{\nu\sigma\rho\mu}}\right)\Bigg),\nonumber 
\end{eqnarray}
and in a compact form it reads 
\begin{equation}
\delta S\left[g_{\mu\nu}\right]=\frac{1}{2}\int_{\mathcal{M}}d\thinspace^{n}x\sqrt{-g}\delta g_{\mu\nu}\biggl(\frac{1}{2}g^{\mu\nu}f\left(R_{\lambda\gamma\rho\sigma}\right)+R^{(\mu}{}_{\gamma\rho\sigma}\frac{\partial f}{\partial R_{\nu)}{}_{\gamma\rho\sigma}}+2\nabla_{\sigma}\nabla_{\rho}\frac{\partial f}{\partial R_{\sigma(\mu\nu)\rho}}\biggr),
\end{equation}
which yields the field equations 
\begin{equation}
-\frac{1}{2}g^{\mu\nu}f\left(R_{\lambda\gamma\rho\sigma}\right)-R^{(\mu}{}_{\gamma\rho\sigma}\frac{\partial f}{\partial R_{\nu)\gamma\rho\sigma}}-2\nabla_{\sigma}\nabla_{\rho}\frac{\partial f}{\partial R_{\sigma(\mu\nu)\rho}}=T_{\mu\nu},\label{Fieldeq}
\end{equation}
where 
\begin{equation}
T_{\mu\nu}=-\frac{2}{\sqrt{-g}}\frac{\delta S_{\text{matter }}}{\delta g^{\mu\nu}}.
\end{equation}

\section{ Introducing auxiliary fields}

To turn the field equations (\ref{Fieldeq}) into a set of first-order
differential equations, we start with the augmented action 
\begin{equation}
S\left[g_{\mu\nu},\rho_{\mu\nu\rho\sigma},\varphi^{\mu\nu\rho\sigma}\right]=\frac{1}{2}\int_{\mathcal{M}}d\thinspace^{n}x\sqrt{-g}\biggl(f\left(\rho_{\mu\nu\rho\sigma}\right)+\varphi^{\mu\nu\rho\sigma}\left(R_{\mu\nu\rho\sigma}-\rho_{\mu\nu\rho\sigma}\right)\biggr),
\end{equation}
where the two auxiliary fields $\rho_{\mu v\rho\sigma}$ and $\varphi^{\mu v\rho\sigma}$
have all the algebraic symmetries of the Riemann tensor $R_{\mu v\rho\sigma}$.

Assuming that the matter couples minimally to the metric and not to
the auxiliary fields, the variation of the action reads

\begin{eqnarray}
\delta S[g,\rho,\varphi] & = & \frac{1}{2}\int_{\mathcal{M}}d\thinspace^{n}x\Big(\delta\sqrt{-g}\biggl(f\left(\rho_{\mu\nu\rho\sigma}\right)+\varphi^{\mu\nu\rho\sigma}\left(R_{\mu\nu\rho\sigma}-\rho_{\mu\nu\rho\sigma}\right)\biggr)\label{varaction}\\
 &  & +\sqrt{-g}\biggl(\delta f\left(\rho_{\mu\nu\rho\sigma}\right)+\delta\varphi^{\mu\nu\rho\sigma}\left(R_{\mu\nu\rho\sigma}-\rho_{\mu\nu\rho\sigma}\right)+\varphi^{\mu\nu\sigma}\left(\delta R_{\mu\nu\rho\sigma}-\delta\rho_{\mu\nu\rho\sigma}\right)\Big)\biggr),\nonumber 
\end{eqnarray}
where 
\begin{eqnarray}
\delta R_{\mu\nu\rho\sigma} & = & R^{\lambda}{}_{\nu\rho\sigma}\delta g_{\mu\lambda}+\frac{1}{2}\biggl(\nabla_{\rho}\nabla_{\nu}\delta g_{\sigma\mu}+\nabla_{\rho}\nabla_{\sigma}\delta g_{\nu\mu}\nonumber \\
 &  & -\nabla_{\rho}\nabla_{\mu}\delta g_{\nu\sigma}-\nabla_{\sigma}\nabla_{\nu}\delta g_{\rho\mu}-\nabla_{\sigma}\nabla_{\rho}\delta g_{\nu\mu}+\nabla_{\sigma}\nabla_{\mu}\delta g_{\nu\rho}\biggr).
\end{eqnarray}
Using the symmetries of the fields, we have 
\begin{eqnarray}
 & \varphi^{\mu\nu\rho\sigma}\left(\delta R_{\mu\nu\rho\sigma}-\delta\rho_{\mu\nu\rho\sigma}\right) & =\varphi^{\mu\nu\rho\sigma}R^{\lambda}\thinspace_{\nu\rho\sigma}\delta g_{\mu\lambda}-\varphi^{\mu\nu\rho\sigma}\delta\rho_{\mu\nu\rho\sigma}\\
 &  & +\frac{1}{2}\varphi^{\mu\nu\rho\sigma}\left(\nabla_{\rho}\nabla_{\nu}\delta g_{\sigma\mu}-\nabla_{\rho}\nabla_{\mu}\delta g_{\nu\sigma}\right.\left.-\nabla_{\sigma}\nabla_{\nu}\delta g_{\rho\mu}+\nabla_{\sigma}\nabla_{\mu}\delta g_{\nu\rho}\right),\nonumber 
\end{eqnarray}
which can be rewritten as

\begin{eqnarray}
\varphi^{\mu\nu\rho\sigma}\left(\delta R_{\mu\nu\rho\sigma}-\delta\rho_{\mu\nu\rho\sigma}\right)=\delta g_{\mu\nu}R^{(\mu}{}_{\lambda\rho\sigma}\varphi^{\nu)\lambda\rho\sigma}+2\varphi^{\sigma(\mu\nu)\rho}\nabla_{\rho}\nabla_{\sigma}\delta g_{\mu\nu}-\varphi^{\mu\nu\rho\sigma}\delta\rho_{\mu\nu\rho\sigma}.
\end{eqnarray}
Inserting these results in (\ref{varaction}), and integrating by
parts, one ends up with 
\begin{eqnarray}
\delta S[g,\rho,\varphi] & = & \frac{1}{2}\int_{\mathcal{M}}d\thinspace^{n}x\sqrt{-g}\Biggl(\delta g_{\mu\nu}\biggl[R^{(\mu}{}_{\lambda\rho\sigma}\varphi^{\nu)\lambda\rho\sigma}+2\nabla_{\sigma}\nabla_{\rho}\varphi^{\sigma(\mu\nu)\rho}\\
 &  & +\frac{1}{2}g^{\mu\nu}\left(f\left(\rho_{\lambda\gamma\rho\sigma}\right)+\varphi^{\lambda\gamma\rho\sigma}\left(R_{\lambda\gamma\rho\sigma}-\rho_{\lambda\gamma\rho\sigma}\right)\right)\biggr]\nonumber \\
 &  & +\left(\frac{\partial f}{\partial\rho_{\mu\nu\rho\sigma}}-\varphi^{\mu\nu\rho\sigma}\right)\delta\rho_{\mu\nu\rho\sigma}+\left(R_{\mu\nu\rho\sigma}-\rho_{\mu\nu\rho\sigma}\right)\delta\varphi^{\mu\nu\rho\sigma}\Biggr)+I_{\text{Boundary}},
\end{eqnarray}
where the boundary terms read 
\begin{eqnarray}
I_{\text{Boundary}}=\int_{\mathcal{M}}d\thinspace^{n}x\sqrt{-g}\left(\nabla_{\rho}\left(\varphi^{\sigma(\mu\nu)\rho}\nabla_{\sigma}\delta g_{\mu\nu}\right)-\nabla_{\sigma}\left(\delta g_{\mu\nu}\nabla_{\rho}\varphi^{\sigma(\mu\nu)\rho}\right)\right).
\end{eqnarray}
Introducing 
\begin{equation}
\mathcal{E^{\mu\nu}}:=-R_{~~\lambda\rho\sigma}^{(\mu}\varphi^{\nu)\lambda\rho\sigma}-2\nabla_{\sigma}\nabla_{\rho}\varphi^{\sigma(\mu\nu)\rho}-\frac{1}{2}g^{\mu\nu}\biggl(f\left(\rho_{\lambda\gamma\rho\sigma}\right)+\varphi^{\lambda\gamma\rho\sigma}\left(R_{\lambda\gamma\rho\sigma}-\rho_{\lambda\gamma\rho\sigma}\right)\biggr),
\end{equation}
and dropping the boundary terms, one arrives at

\begin{equation}
\delta S=\frac{1}{2}\int d\thinspace^{n}x\sqrt{-g}\left(-\mathcal{E^{\mu\nu}}\delta g_{\mu\nu}+\left(R_{\mu\nu\rho\sigma^{-}}\rho_{\mu\nu\rho\sigma}\right)\delta\varphi^{\mu\nu\rho\sigma}+\left(\frac{\partial f}{\partial\rho_{\mu\nu\rho\sigma}}-\varphi^{\mu\nu\rho\sigma}\right)\delta\rho_{\mu\nu\rho\sigma}\right).
\end{equation}
The field equations given in Section II follow from the above variation.
One can show that using the field equations for the auxiliary fields
in the $\mathcal{E^{\mu\nu}}=T^{\mu\nu}$ equation, one gets back
the correct second-order field equations (\ref{EOM1}), hence the
consistency.

\section{ADM splitting of the auxiliary fields}

Let us give some details of the computations leading to the action
(\ref{red}). One has 

\begin{eqnarray}
\varphi^{\mu\nu\rho\sigma}\left(R_{\mu\nu\rho\sigma}-\rho_{\mu\nu\rho\sigma}\right) & = & \varphi^{0\nu\rho\sigma}\left(R_{0\nu\rho\sigma}-\rho_{0\nu\rho\sigma}\right)+\varphi^{i\nu\rho\sigma}\left(R_{i\nu\rho\sigma}-\rho_{i\nu\rho\sigma}\right)\\
 & = & \varphi^{0i\rho\sigma}\left(R_{0i\rho\sigma}-\rho_{0i\rho\sigma}\right)+\varphi^{i0\rho\sigma}\left(R_{i0\rho\sigma}-\rho_{i0\rho\sigma}\right)+\varphi^{ij\rho\sigma}\left(R_{ij\rho\sigma}-\rho_{ij\rho\sigma}\right).\nonumber 
\end{eqnarray}
Due to the symmetries, it can be written as 
\begin{equation}
\varphi^{\mu\nu\rho\sigma}\left(R_{\mu\nu\rho\sigma}-\rho_{\mu\nu\rho\sigma}\right)=2\varphi^{0i\rho\sigma}\left(R_{0i\rho\sigma}-\rho_{0i\rho\sigma}\right)+\varphi^{ij\rho\sigma}\left(R_{ij\rho\sigma}-\rho_{ij\rho\sigma}\right),
\end{equation}
which yields 
\begin{align}
\varphi^{\mu\nu\rho\sigma}\left(R_{\mu\nu\rho\sigma}-\rho_{\mu\nu\rho\sigma}\right) & =2\varphi^{0i0\sigma}\left(R_{0i0\sigma}-\rho_{0i0\sigma}\right)+2\varphi^{0ij\sigma}\left(R_{0ij\sigma}-\rho_{0ij\sigma}\right)\nonumber \\
 & +\varphi^{ij0\sigma}\left(R_{ij0\sigma}-\rho_{ij0\sigma}\right)+\varphi^{ijk\sigma}\left(R_{ijk\sigma}-\rho_{ijk\sigma}\right),
\end{align}
and then 
\begin{align}
\varphi^{\mu\nu\rho\sigma}\left(R_{\mu\nu\rho\sigma}-\rho_{\mu\nu\rho\sigma}\right) & =2\varphi^{0i0j}\left(R_{0i0j}-\rho_{0i0j}\right)+2\varphi^{0ij0}\left(R_{0ij0}-\rho_{0ij0}\right)+2\varphi^{0ijk}\left(R_{0ijk}-\rho_{0ijk}\right)\nonumber \\
 & +\varphi^{ij0k}\left(R_{ij0k}-\rho_{ij0k}\right)+\varphi^{ijk0}\left(R_{ijk0}-\rho_{ijk0}\right)+\varphi^{ijkl}\left(R_{ijkl}-\rho_{ijkl}\right).
\end{align}
Using the symmetries, one ends up with 
\begin{eqnarray*}
\varphi^{\mu\nu\rho\sigma}\left(R_{\mu\nu\rho\sigma}-\rho_{\mu\nu\rho\sigma}\right)=\varphi^{ijkl}\left(R_{ijkl}-\rho_{ijkl}\right)+4\varphi^{ijk0}\left(R_{ijk0}-\rho_{ijk0}\right)+4\varphi^{i0j0}\left(R_{i0j0}-\rho_{i0j0}\right).
\end{eqnarray*}
Moreover, using the decomposition of the components of the Riemann
tensor one obtains 
\begin{align}
\varphi^{\mu\nu\rho\sigma}\left(R_{\mu\nu\rho\sigma}-\rho_{\mu\nu\rho\sigma}\right)= & \varphi^{ijkl}\left(R_{ijkl}-\rho_{ijkl}\right)\\
 & +4\varphi^{ijk0}\left(N\left(D_{i}K_{jk}-D_{j}K_{ik}\right)+N^{l}R_{ijkl}-\rho_{ijk0}\right)\nonumber \\
 & +4\varphi^{i0j0}\biggl(NN^{k}\left(D_{i}K_{jk}+D_{j}K_{ik}-2D_{k}K_{ij}\right)\nonumber \\
 & +N\left(-\dot{K}_{ij}+\mathcal{L}_{N}K_{ij}+D_{i}D_{j}N\right)+N^{2}K_{ik}K^{k}{}_{j}+N^{k}N^{l}R_{ikjl}-\rho_{i0j0}\biggr).\nonumber 
\end{align}
Defining 
\begin{eqnarray}
 &  & \varphi^{ijk}\equiv\gamma^{il}\gamma^{jm}\gamma^{kn}n^{\mu}\varphi_{lmn\mu},\hskip1cm\psi^{ij}\equiv-2\gamma^{ik}\gamma^{jl}n^{\mu}n^{v}\varphi_{k\mu l\nu},\nonumber \\
 &  & \rho_{ijk}\equiv n^{\mu}\rho_{ijk\mu,}\hskip3cm\Omega_{ij}\equiv n^{\mu}n^{\nu}\rho_{i\mu j\nu},
\end{eqnarray}
one can express the corresponding components of the auxiliary field
$\varphi^{\mu\nu\rho\sigma}$ in terms of the spatial tensor fields
as follows 
\begin{equation}
\varphi^{ijk0}=-\frac{\varphi^{ijk}}{N},~~~~~~~~~~~~~~~~\varphi^{i0j0}=-\frac{\psi^{ij}}{2N^{2}},
\end{equation}
and similarly, in terms of the spatial tensors, we can write the components
of $\rho_{\mu\nu\rho\sigma}$ as 
\begin{align}
\rho_{ijk0} & =N\rho_{ijk}+N^{l}\rho_{ijkl},\\
\rho_{ijk0} & =N^{2}\Omega_{ij}+N^{k}N^{l}\rho_{ikjl}+NN^{k}\left(\rho_{ikj}+\rho_{jki}\right).\nonumber 
\end{align}
Then, we arrive at

\begin{align}
\varphi^{\mu\nu\rho\sigma}\left(R_{\mu\nu\rho\sigma}-\rho_{\mu\nu\rho\sigma}\right)= & \varphi^{ijkl}\left(R_{ijkl}-\rho_{ijkl}\right)\nonumber \\
 & -4\varphi^{ijk}\left(D_{i}K_{jk}-D_{j}K_{ik}-\rho_{ijk}+\frac{N^{l}}{N}\left(R_{ijkl}-\rho_{ijkl}\right)\right)\nonumber \\
 & -2\psi^{ij}\biggl(\frac{N^{k}N^{l}}{N^{2}}\left(R_{ikjl}-\rho_{ikjl}\right)+K_{ik}K^{k}{}_{j}-\Omega_{ij}\nonumber \\
 & \quad+\frac{N^{k}}{N}\left(D_{i}K_{jk}+D_{j}K_{ik}-2D_{k}K_{ij}-\rho_{ikj}-\rho_{jki}\right)\nonumber \\
 & +\frac{1}{N}\left(-\dot{K_{ij}}+\mathcal{L}_{N}K_{ij}+D_{i}D_{j}N\right)\biggr),
\end{align}
from which one obtains (\ref{red}).

\section{The constraints}

By definition, the canonical momenta are defined as 
\begin{equation}
\Pi_{ij}:=\frac{\delta\mathcal{L}}{\delta\partial_{0}\psi^{ij}},~~~~~~~~~~~~~~~~~p^{ij}:=\frac{\delta\mathcal{L}}{\delta\partial_{0}\gamma_{ij}}=\frac{\delta\mathcal{L}}{\delta K_{lm}}\frac{\delta K_{lm}}{\delta\dot{\gamma}_{ij}},
\end{equation}
where the Lagrangian density is given in (\ref{lagrangian}), and
so we can express 
\begin{equation}
\Pi_{ij}=-\sqrt{\gamma}K_{ij}.
\end{equation}
In the computation of $p^{ij}$, one needs to compute the variation
of the extrinsic curvature with respect to $\dot{\gamma}_{ij}$. By
definition, we have 
\begin{equation}
\frac{\delta K_{lm}}{\delta\dot{\gamma}_{ij}}=\frac{1}{2N}\frac{\delta}{\delta\dot{\gamma}_{ij}}\left(\dot{\gamma}_{lm}-D_{l}N_{m}-D_{m}N_{l}\right)=\frac{1}{2N}\delta_{l}^{i}\delta_{m}^{j}.
\end{equation}
One also uses the explicit form of the Lagrangian (\ref{lagrangian})
to arrive at 
\begin{eqnarray*}
\frac{\delta\mathcal{L}}{\delta K_{lm}} & = & \frac{N\sqrt{\gamma}}{2}\frac{\delta f}{\delta K_{lm}}+N\sqrt{\gamma}\left(\frac{1}{N}\left(\mathcal{L}_{N}\psi^{lm}-\dot{\psi}^{lm}\right)-\gamma^{lm}\psi^{ps}K_{ps}-K\psi^{lm}-\psi^{ln}K_{n}^{m}-\psi^{mn}K_{n}^{l}\right).
\end{eqnarray*}
Collecting the pieces we express the conjugate momenta as 
\begin{align*}
 & p^{ij}=\frac{\sqrt{\gamma}}{4}\frac{\delta f}{\delta K_{ij}}+\frac{\sqrt{\gamma}}{2}\left(\frac{1}{N}\left(\mathcal{L}_{N}\psi^{ij}-\dot{\psi}^{ij}\right)-\gamma^{ij}\psi^{kl}K_{kl}-K\psi^{ij}-\psi^{ik}K_{k}{}^{j}-\psi^{jk}K_{k}^{i}\right).
\end{align*}
The last expression directly yields the velocities as 
\begin{align}
\dot{\psi}^{ij} & =\frac{N}{2}\frac{\delta f}{\delta K_{ij}}-\frac{2N}{\sqrt{\gamma}}p^{ij}+\mathcal{L_{N}}\psi^{ij}+N\left(-\gamma^{ij}\psi^{kl}K_{kl}-K\psi^{ij}-\psi^{ik}K_{k}^{j}-\psi^{jk}K_{k}^{i}\right).
\end{align}
The Hamiltonian density reads

\begin{equation}
\mathcal{H}=p^{ij}\dot{\gamma}_{ij}+\Pi_{ij}\dot{\psi}^{ij}-\mathcal{L},
\end{equation}
and inserting the velocities together with the Lagrangian density
(\ref{lagrangian}) it can be expressed as follows

\begin{align}
\mathcal{H}= & 2NK_{ij}~p^{ij}-\sqrt{\gamma}K_{ij}\mathcal{L}_{N}\psi^{ij}+2p^{ij}D_{i}N_{j}\\
 & +\sqrt{\gamma}N\left(-\frac{f}{2}+\frac{1}{N}\psi^{ij}D_{i}D_{j}N-\psi^{ij}\Omega_{ij}+K\thinspace\psi^{kl}K_{kl}+K_{ij}\psi^{ik}K_{k}^{j}\right).\nonumber 
\end{align}
Up to a boundary term, the Hamiltonian density can be written as a
sum of the constraint equations. Namely, one has

\begin{equation}
\mathcal{H}=N\Phi_{0}+N^{i}\Phi_{i}.
\end{equation}
After a straightforward computation, one obtains

\begin{align}
\mathcal{H}= & N\sqrt{\gamma}\left(\frac{2}{\sqrt{\gamma}}K_{ij}\thinspace p{}^{ij}-\frac{1}{2}f\left(\rho_{\mu v\rho\sigma}\right)+D_{i}D_{j}\psi^{ij}-\psi^{ij}\Omega_{ij}+K\psi^{kl}K_{kl}+K_{ij}K_{k}^{j}\psi^{ik}\right)\nonumber \\
 & +N^{i}\sqrt{\gamma}\left(-2D_{k}\left(\frac{p_{i}^{k}}{\sqrt{\gamma}}\right)-K_{kl}D_{i}\psi^{kl}-2D_{k}\left(\psi^{kl}K_{li}\right)\right).
\end{align}
Since we have already obtained the relation between the extrinsic
curvature and the conjugate momenta, $K_{ij}=-\Pi_{ij}/\sqrt{\gamma}$,
we can equivalently write

\begin{align}
\mathcal{H}= & N\sqrt{\gamma}\left(D_{i}D_{j}\psi^{ij}-\psi^{ij}\thinspace\Omega_{ij}-\frac{1}{2}f(\rho_{\mu v\rho\sigma}) \right)+\frac{N}{\sqrt{\gamma}}\left(-2\Pi_{ij}p^{ij}+\Pi\thinspace\Pi_{ij}\thinspace\psi^{ij}+\Pi_{ij}\Pi_{k}^{j}\psi^{ik}\right)\nonumber \\
 & +N^{i}\left(-2\sqrt{\gamma}D_{k}\left(\frac{p_{i}^{k}}{\sqrt{\gamma}}\right)+\Pi_{kl}D_{i}\psi^{kl}+2\sqrt{\gamma}D_{k}\left(\psi^{kl}\frac{\Pi_{li}}{\sqrt{\gamma}}\right)\right),
\end{align}
which yields the Hamiltonian and the momentum constraints as

\begin{align}
\Phi_{0}= & \sqrt{\gamma}\left(D_{i}D_{j}\psi^{ij}-\psi^{ij}\thinspace\Omega_{ij}-\frac{1}{2}f(\rho_{\mu v\rho\sigma})\right)+\frac{1}{\sqrt{\gamma}}\left(-2\Pi_{ij}\thinspace p^{ij}+\Pi_{ij}\thinspace\psi^{ij}+\Pi_{ij}\Pi_{k}^{j}\psi^{ik}\right),\\
\Phi_{i}= & -2\sqrt{\gamma}D_{k}\left(\frac{p_{i}^{k}}{\sqrt{\gamma}}\right)+\Pi_{kl}D_{i}\psi^{kl}+2\sqrt{\gamma}D_{k}\left(\psi^{kl}\frac{\Pi_{li}}{\sqrt{\gamma}}\right).
\end{align}

\section{Time evolution equations}

As for the dynamical equations, the first set of the evolution equations
is

\begin{equation}
\dot{\gamma}_{ij}=\frac{\delta\mathcal{H}}{\delta p^{ij}},\quad\quad\quad\quad\quad\quad\quad\dot{\psi}^{ij}=\frac{\delta\mathcal{H}}{\delta\Pi_{ij}}.
\end{equation}
This set gives the velocities in terms of canonical variables. Ignoring
the total derivative terms, the Hamiltonian density can be expressed
as

\begin{equation}
\begin{aligned}\mathcal{H}= & N\sqrt{\gamma}\left(D_{k}D_{l}\psi^{kl}-\psi^{kl}\Omega_{kl}-\frac{1}{2}f(\rho_{\mu v\rho\sigma})\right)+\frac{N}{\sqrt{\gamma}}\left(-2\Pi_{kl}p^{kl}+\Pi~\Pi_{kl}\psi^{kl}+\Pi_{kl}\Pi_{m}^{k}\psi^{ml}\right)\\
 & +N^{m}\left(-2\sqrt{\gamma}D_{k}\left(\frac{p_{m}^{k}}{\sqrt{\gamma}}\right)+\Pi_{kl}D_{m}\psi^{kl}+2\sqrt{\gamma}D_{k}\left(\psi{}^{kl}\frac{\Pi_{lm}}{\sqrt{\gamma}}\right)\right).
\end{aligned}
\end{equation}
Then automatically we obtain 
\begin{equation}
\dot{\gamma}_{ij}=-\frac{2N}{\sqrt{\gamma}}\Pi_{ij}+\mathcal{L}_{N}\gamma_{ij}=2NK_{ij}+D_{i}N_{j}+D_{j}N_{i},
\end{equation}
which is the expected result (\ref{evolution=000020spatial=000020metric})
and by definition Lie derivative yields $\mathcal{L}_{N}\gamma_{ij}=D_{i}N_{j}+D_{j}N_{i}$.
Similarly $\dot{\psi}^{ij}$ can be written as 
\begin{equation}
\dot{\psi}^{ij}=\frac{N}{\sqrt{\gamma}}\left(-2p^{ij}+\gamma^{ij}\Pi_{kl}\psi^{kl}+\Pi\thinspace\psi^{ij}+\Pi_{k}^{i}\psi^{kj}+\Pi_{k}^{j}\psi^{ki}\right)+\mathcal{L}_{N}\psi^{ij}-\frac{N\sqrt{\gamma}}{2}\frac{\delta f}{\delta\Pi_{ij}},
\end{equation}
where one can evaluate the last term easily for a given $f$.

Now, we can continue with the second set of the evolution equations.
One has the following relations

\begin{equation}
\dot{p}^{ij}=-\frac{\delta\mathcal{H}}{\delta\gamma_{ij}},\quad\quad\quad\quad\quad\quad\quad\dot{\Pi}_{ij}=-\frac{\delta\mathcal{H}}{\delta\psi^{ij}}.
\end{equation}
Using the Hamiltonian density again let us construct $\dot{\Pi}_{ij}$.
We have

\begin{equation}
\begin{aligned}\dot{\Pi}_{ij}= & N\sqrt{\gamma}~\Omega_{ij}-\frac{N}{\sqrt{\gamma}}\Pi~\Pi_{ij}-\frac{N}{\sqrt{\gamma}}\Pi_{ik}\Pi^{k}{}_{j}-N\sqrt{\gamma}\frac{\delta}{\delta\psi^{ij}}\left(D_{k}D_{1}\psi^{kl}\right)\\
 & -N^{m}\Pi_{kl}\frac{\delta}{\delta\psi^{ij}}\left(D_{m}\psi^{kl}\right)-2N^{m}\frac{\delta}{\delta\psi^{ij}}\left(D_{k}(\psi^{kl}\Pi_{lm})\right),
\end{aligned}
\end{equation}
and we have to compute the last three terms. We have

\begin{equation}
D_{m}\psi^{kl}=\partial_{m}\psi^{kl}+\thinspace^{\Sigma}\Gamma_{mn}^{k}\psi^{nl}+\thinspace^{\Sigma}\Gamma_{mn}^{l}\psi^{kn},
\end{equation}
which yields the following variation 
\begin{equation}
\begin{array}{r}
\delta D_{m}\psi^{kl}=\partial_{m}\delta\psi^{kl}+{}^{\Sigma}\Gamma_{mn}^{k}\delta\psi^{n1}+\thinspace^{\Sigma}\Gamma_{mn}^{l}\delta\psi^{kn}+\psi^{nl}\delta\thinspace^{\Sigma}\Gamma_{mn}^{k}+\psi^{kn}\delta\thinspace^{\Sigma}\Gamma_{mn}^{l}\end{array},
\end{equation}
and it can be written in a more compact form as 
\begin{equation}
\begin{array}{r}
\delta D_{m}\psi^{kl}=D_{m}\delta\psi^{kl}+\psi^{nl}\delta\thinspace^{\Sigma}\Gamma_{mn}^{k}+\psi^{kn}\delta\thinspace^{\Sigma}\Gamma_{mn}^{l}\end{array}.
\end{equation}
The variation of the hypersurface connection can be expressed as

\begin{equation}
\delta{}^{\Sigma}\Gamma_{mn}^{k}=\frac{1}{2}\gamma^{kp}\left(D_{m}\delta\gamma_{np}+D_{n}\delta\gamma_{mp}-D_{p}\delta\gamma_{mn}\right).\label{variation=000020hypersurface=000020connection}
\end{equation}
Therefore in a more explicit form, one obtains

\begin{equation}
\begin{aligned}\delta D_{m}\psi^{kl}= & D_{m}\delta\psi^{kl}+\frac{1}{2}\psi^{nl}\gamma^{kp}\left(D_{m}\delta\gamma_{np}+D_{n}\delta\gamma_{mp}-D_{p}\delta\gamma_{mn}\right)\\
 & +\frac{1}{2}\psi^{kn}\gamma^{1p}\left(D_{m}\delta\gamma_{np}+D_{n}\delta\gamma_{mp}-D_{p}\delta\gamma_{mn}\right),
\end{aligned}
\end{equation}
which can be reexpressed as 
\begin{equation}
\delta D_{m}\psi^{kl}=D_{m}\delta\psi^{kl}+\psi^{n(k}\gamma^{l)p}\left(D_{m}\delta\gamma_{np}+D_{n}\delta\gamma_{mp}-D_{p}\delta\gamma_{mn}\right),
\end{equation}
and directly yields variation with respect to $\psi^{ij}$ as

\[
N^{m}\Pi_{kl}\frac{\delta\left(D_{m}\psi^{kl}\right)}{\delta\psi^{ij}}=-N^{m}D_{m}\Pi_{ij}-\Pi_{ij}D_{m}N^{m}.\tag{68}
\]
Note that there is no contribution coming from the variations of the
connection in the last expression. Similarly we can compute $\delta D_{k}D_{l}\psi^{kl}$
as

\begin{equation}
\begin{aligned}D_{k}D_{l}\psi^{kl}= & \partial_{k}\left(\partial_{l}\psi^{kl}+\thinspace^{\Sigma}\Gamma_{lm}^{k}\psi^{ml}+\thinspace^{\Sigma}\Gamma_{lm}^{l}\psi^{km}\right)+\thinspace^{\Sigma}\Gamma_{km}^{k}\left(\partial_{l}\psi^{ml}+\thinspace^{\Sigma}\Gamma_{ln}^{m}\psi^{nl}+\thinspace^{\Sigma}\Gamma_{ln}^{l}\psi^{mn}\right).\end{aligned}
\end{equation}
Taking the variation, we write

\begin{align}
\delta D_{k}D_{l}\psi^{kl} & =D_{k}D_{l}\delta\psi^{kl}+D^{p}\left(\psi^{ml}D_{l}\delta\gamma_{mp}\right)-\frac{1}{2}D^{k}\left(\psi^{ml}D_{k}\delta\gamma_{lm}\right)\nonumber \\
 & +\frac{1}{2}D_{k}\left(\psi^{km}\gamma^{lp}D_{m}\delta\gamma_{lp}\right)+\frac{1}{2}\gamma^{kp}D_{l}\psi^{ml}D_{m}\delta\gamma_{kp}.
\end{align}
Since we focus on variation of the spatial field $\psi$, we only
consider the first term on the right-hand side. Hence, ignoring the
total derivative terms, we get

\begin{equation}
N\sqrt{\gamma}\frac{\delta}{\delta\psi^{ij}}\left(D_{k}D_{l}\psi^{kl}\right)=\sqrt{\gamma}D_{i}D_{j}N.
\end{equation}
Now we should compute the last term: $\delta D_{k}\left(\psi^{kl}\Pi_{1m}\right)$.
The variation of this term gives us

\[
\delta D_{k}\left(\psi^{kl}\Pi_{lm}\right)=D_{k}\delta\left(\psi^{kl}\Pi_{lm}\right)-\psi^{kl}\Pi_{ln}\delta^{\Sigma}\Gamma_{km}^{n}.\tag{71}
\]
Up to a boundary term we get

\[
N^{m}\frac{\delta}{\delta\psi^{ij}}D_{k}\left(\psi^{kl}\Pi_{lm}\right)=-\Pi_{m(i}D_{j)}N^{m}.\tag{72}
\]
Inserting them in $\dot{\Pi}_{ij}$, we end up with the desiring evolution
equation

\begin{equation}
\dot{\Pi}_{ij}=\sqrt{\gamma}\left(N\Omega_{ij}-D_{i}D_{j}N\right)-\frac{N}{\sqrt{\gamma}}\left(\Pi\thinspace\Pi_{ij}+\Pi_{ik}\Pi_{j}^{k}\right)+\mathcal{L}_{N}\Pi_{ij}+\Pi_{ij}D_{k}N^{k}.
\end{equation}
Then, we can construct $\dot{p}^{ij}=-\delta\mathcal{H}/\delta\gamma_{ij}$.
Since we have the variations 
\begin{equation}
\frac{\delta\sqrt{\gamma}}{\delta\gamma_{ij}}=\frac{1}{2}\sqrt{\gamma}\gamma^{ij},~~~~~~~~~~~~~~~~~~~~~~\frac{\delta\gamma^{-1/2}}{\delta\gamma_{ij}}=-\frac{1}{2}\gamma^{-1/2}\gamma^{ij},
\end{equation}
from the Hamiltonian density (\ref{Hamiltonain}) we directly obtain
\begin{align}
\dot{p}^{ij} & =-\frac{N}{2}\sqrt{\gamma}\gamma^{ij}\left(D_{k}D_{l}\psi^{kl}-\psi^{kl}\Omega_{kl}-\frac{1}{2}f(\rho_{\mu v\rho\sigma})\right)-N\sqrt{\gamma}\frac{\delta}{\delta\gamma_{ij}}\left(D_{k}D_{l}\psi^{kl}-\frac{1}{2}f(\rho_{\mu v\rho\sigma})\right)\nonumber \\
 & +\frac{N}{2\sqrt{\gamma}}\gamma^{ij}\left(-2\Pi_{kl}p^{kl}+\Pi~\Pi_{kl}\psi^{kl}+\Pi_{kl}\Pi_{m}^{k}\psi^{ml}\right)\nonumber \\
 & -\frac{N}{\sqrt{\gamma}}\frac{\delta}{\delta\gamma_{ij}}\left(\Pi_{mn}\gamma^{mn}~\Pi_{kl}\psi^{kl}+\Pi_{kl}\Pi_{mn}\gamma^{kn}\psi^{ml}\right)\nonumber \\
 & -N^{m}\frac{\delta}{\delta\gamma_{ij}}\left(-2\gamma_{mn}\sqrt{\gamma}D_{k}\left(\frac{p^{kn}}{\sqrt{\gamma}}\right)+\Pi_{kl}D_{m}\psi^{kl}+2\sqrt{\gamma}D_{k}\left(\psi{}^{kl}\frac{\Pi_{lm}}{\sqrt{\gamma}}\right)\right).\label{eq:denklem}
\end{align}
To simplify the last equation, we use the Hamiltonian constraint (\ref{Hamiltonian=000020constraint})
together with 
\begin{equation}
\frac{\delta\gamma^{mn}}{\delta\gamma_{ij}}=-\frac{1}{2}\left(\gamma^{mi}\gamma^{jn}+\gamma^{mj}\gamma^{in}\right),
\end{equation}
in (\ref{eq:denklem}) to arrive at 
\begin{align}
\dot{p}^{ij} & =\frac{N}{\sqrt{\gamma}}\Biggl(\gamma^{ij}\left(\Pi~\Pi_{kl}\psi^{kl}+\Pi_{kl}\Pi_{m}^{k}\psi^{ml}-2\Pi_{kl}p^{kl}\right)+\Pi^{ij}\Pi_{kl}\psi^{kl}+\Pi_{l}^{i}\Pi_{k}^{j}\psi^{kl}\Biggr)\nonumber \\
 & -N\sqrt{\gamma}\frac{\delta}{\delta\gamma_{ij}}\left(D_{k}D_{l}\psi^{kl}\right)+\frac{N}{2}\sqrt{\gamma}\frac{\delta}{\delta\gamma_{ij}}f(\rho_{\mu v\rho\sigma})\nonumber \\
 & -N^{m}\frac{\delta}{\delta\gamma_{ij}}\left(-2\gamma_{mn}\sqrt{\gamma}D_{k}\left(\frac{p^{kn}}{\sqrt{\gamma}}\right)+\Pi_{kl}D_{m}\psi^{kl}+2\sqrt{\gamma}D_{k}\left(\psi{}^{kl}\frac{\Pi_{lm}}{\sqrt{\gamma}}\right)\right).
\end{align}
A straightforward calculation gives us 
\begin{equation}
N\sqrt{\gamma}\frac{\delta}{\delta\gamma_{ij}}\left(D_{k}D_{l}\psi^{kl}\right)=-\frac{\sqrt{\gamma}}{2}\biggl(D_{k}\left(\psi^{ij}D^{k}N-2\psi^{k(i}D^{j)}N\right)+\gamma^{ij}\left(ND_{k}D_{l}\psi^{kl}-\psi^{kl}D_{k}D_{l}N\right)\biggr),
\end{equation}
and 
\begin{equation}
2N^{m}\frac{\delta}{\delta\gamma_{ij}}\left(\gamma_{mn}\sqrt{\gamma}D_{k}\left(\frac{p^{kn}}{\sqrt{\gamma}}\right)\right)=\sqrt{\gamma}\mathcal{L}_{N}\left(\frac{p^{ij}}{\sqrt{\gamma}}\right)+p^{ij}D_{k}N^{k},
\end{equation}
and also 
\begin{equation}
N^{m}\frac{\delta}{\delta\gamma_{ij}}\left(\Pi_{kl}D_{m}\psi^{kl}+2\sqrt{\gamma}D_{k}\left(\psi{}^{kl}\frac{\Pi_{lm}}{\sqrt{\gamma}}\right)\right)=0.
\end{equation}
Finally, after collecting the pieces, one ends up with the last evolution
equation (\ref{evolution=000020pij}) 
\begin{align}
\dot{p}^{ij} & =\frac{N}{\sqrt{\gamma}}\Biggl(\gamma^{ij}\left(\Pi~\Pi_{kl}\psi^{kl}+\Pi_{kl}\Pi_{m}^{k}\psi^{ml}-2\Pi_{kl}p^{kl}\right)+\Pi^{ij}\Pi_{kl}\psi^{kl}+\Pi_{l}^{i}\Pi_{k}^{j}\psi^{kl}\Biggr)\nonumber \\
 & +\frac{\sqrt{\gamma}}{2}\biggl(D_{k}\left(\psi^{ij}D^{k}N-2\psi^{k(i}D^{j)}N\right)+\gamma^{ij}\left(ND_{k}D_{l}\psi^{kl}-\psi^{kl}D_{k}D_{l}N\right)\biggr)\nonumber \\
 & +\frac{N}{2}\sqrt{\gamma}\frac{\delta f}{\delta\gamma_{ij}}+\sqrt{\gamma}\mathcal{L}_{N}\left(\frac{p^{ij}}{\sqrt{\gamma}}\right)+p^{ij}D_{k}N^{k}.
\end{align}
At this point, one needs to know the explicit form of the function $f$ 
to proceed further.

\section{Application to General Relativity}

In this section, we will use the generic results to evaluate Einstein's
theory. Let us set the function $f$ to 
\begin{equation}
f(\rho_{\mu\nu\rho\sigma})=g^{\mu\rho}g^{\nu\sigma}\rho_{\mu\nu\rho\sigma}.
\end{equation}
Since $\rho_{\mu\nu\rho\sigma}$ has the symmetries of the Riemann
tensor by assumption, one can easily show that 
\begin{equation}
f=g^{ik}g^{jl}\rho_{ijkl}+4g^{0j}g^{ik}\rho_{0ijk}+2\rho_{0i0j}\left(g^{00}g^{ij}-g^{0i}g^{0j}\right).
\end{equation}
We can insert the components of the inverse spacetime metric to arrive
at 
\begin{equation}
f(\rho_{\mu\nu\rho\sigma})=\gamma^{ik}\gamma^{jl}\rho_{ijkl}+\frac{2}{N^{2}}\gamma^{ij}\left(2N^{k}\rho_{0ikj}-\rho_{0i0j}-N^{k}N^{l}\rho_{ikjl}\right).
\end{equation}
Using the hypersurface projected tensor fields that we introduced
before, we arrive at 
\begin{equation}
f=\gamma^{ik}\gamma^{jl}\rho_{ijkl}-2\gamma^{ij}\Omega_{ij}.
\end{equation}
Now we can construct the constraint on the auxiliary field $\psi^{ij}$,
(\ref{constraint=000020on=000020psi}). It directly yields 
\begin{equation}
\psi^{ij}=\gamma^{ij}.
\end{equation}
Therefore, the first set of evolution equations 
(\ref{evolution=000020spatial=000020metric},
\ref{evolution=000020psi}) are related. Since $\psi^{ij}=\gamma^{ij}$,
we can rewrite (\ref{evolution=000020psi}) as 
\begin{equation}
\dot{\gamma}^{ij}=-D^{i}N^{j}-D^{j}N^{i}+\frac{2N}{\sqrt{\gamma}}\left(\Pi\gamma^{ij}+\Pi^{ij}-p^{ij}-\Pi_{kl}\frac{\partial f}{\partial\rho_{ikjl}}\right),
\end{equation}
where we can use 
\begin{equation}
\frac{\partial f}{\partial\rho_{ikjl}}=\gamma^{pq}\gamma^{nm}\frac{\partial\rho_{pnqm}}{\partial\rho_{ikjl}}.
\end{equation}
We have to preserve the symmetries on both sides of the equation.
So, we have to express it in a more correct form as 
\begin{equation}
\frac{\partial f}{\partial\rho_{ikjl}}=\frac{1}{4}\gamma^{pq}\gamma^{nm}\frac{\partial}{\partial\rho_{ikjl}}\left(\rho_{pnqm}+\rho_{qmpn}-\rho_{npqm}-\rho_{pnmq}\right),
\end{equation}
which yields 
\begin{equation}
\frac{\partial f}{\partial\rho_{ikjl}}=\frac{1}{2}\left(\gamma^{ij}\gamma^{kl}-\gamma^{il}\gamma^{kj}\right).
\end{equation}
Then we have 
\begin{equation}
\dot{\gamma}^{ij}=-D^{i}N^{j}-D^{j}N^{i}+\frac{2N}{\sqrt{\gamma}}\left(\frac{1}{2}\Pi\gamma^{ij}+\frac{3}{2}\Pi^{ij}-p^{ij}\right),
\end{equation}
Using the basic relation $\dot{\gamma}_{ij}=-\gamma_{ik}\gamma_{jl}\dot{\gamma}^{kl}$,
one can directly rewrite the last expression as 
\begin{equation}
\dot{\gamma}_{ij}=D_{i}N_{j}+D_{j}N_{i}+\frac{2N}{\sqrt{\gamma}}\left(-\frac{1}{2}\Pi\gamma_{ij}+p_{ij}-\frac{3}{2}\Pi_{ij}\right).
\end{equation}
For consistency with the time evolution of the spatial metric 
\begin{equation}
\dot{\gamma}_{ij}=-\frac{2N}{\sqrt{\gamma}}\Pi_{ij}+D_{i}N_{j}+D_{j}N_{i},
\end{equation}
one needs 
\begin{equation}
\Pi_{ij}=2\left(p_{ij}-\frac{p}{n}\gamma_{ij}\right).\label{relation}
\end{equation}

\subsection{Constraint equations}

One can reexpress the Hamiltonian density of General Relativity, using
$\Psi^{ij}=\gamma^{ij}$ and (\ref{relation}) as 
\begin{equation}
\mathcal{H}=\left(-p^{ij}+\frac{2p}{n}\gamma^{ij}\right)\dot{\gamma}_{ij}-\mathcal{L}.
\end{equation}
We then introduce the new momentum 
\begin{equation}
\mathcal{\mathcal{\pi}}^{ij}:=-p^{ij}+\frac{2p}{n}\gamma^{ij},
\end{equation}
which yields the trace 
\begin{equation}
\pi=\frac{n-2}{n}p.
\end{equation}
Then, the Hamiltonian density becomes 
\begin{equation}
\mathcal{H}=\mathcal{\mathcal{\mathcal{\mathcal{\pi}}}}^{ij}\dot{\gamma}_{ij}-\mathcal{L}.
\end{equation}
One has the inverse relations 
\begin{equation}
\Pi^{ij}=-2\mathcal{\mathcal{\mathcal{\mathcal{\pi}}}}^{ij}+\frac{2}{n-2}\gamma^{ij}\mathcal{\mathcal{\mathcal{\mathcal{\pi}}}},
\end{equation}
and $\Pi=2\pi/(n-2)$. Additionally, we can express 
\begin{equation}
p^{ij}=-\mathcal{\mathcal{\mathcal{\mathcal{\pi}}}}^{ij}+\frac{2}{n-2}\gamma^{ij}\pi,~~~~~~~~~~~~p=\frac{n}{n-2}\mathcal{\mathcal{\mathcal{\mathcal{\pi}}}}.
\end{equation}
Finally, the Hamiltonian and the momentum constraint equations (\ref{Hamiltonian=000020constraint},
\ref{momentum=000020constraint}) reduce to 
\begin{equation}
\Phi_{0}(\gamma,\pi)=\frac{2}{\sqrt{\gamma}}(\mathcal{\mathcal{\mathcal{\mathcal{\pi}}}}_{ij}^{2}-\frac{\mathcal{\mathcal{\mathcal{\mathcal{\pi}}}}^{2}}{n-2})-\frac{\sqrt{\gamma}}{2}~{}^{\Sigma}R,
\end{equation}
and 
\begin{equation}
\Phi_{i}(\gamma,\pi)=-2\mathcal{D}_{k}\mathcal{\mathcal{\mathcal{\mathcal{\pi}}}}{}_{i}^{k}.
\end{equation}

\end{document}